\documentclass[aps,reprint]{revtex4-1}
\usepackage{graphicx}
\usepackage{enumitem}
\usepackage{amsmath}
\newcommand{\ket}[1]{|{#1}\rangle}
\newcommand{\bra}[1]{\langle{#1}|}
\usepackage{amsfonts}

\begin{document}
\title{Driven Topological Systems in the Classical Limit}
\author{Callum W. Duncan}
\email{cd130@hw.ac.uk}
\author{Patrik \"{O}hberg}
\author{Manuel Valiente}
\affiliation{SUPA, Institute of Photonics and Quantum Sciences,
Heriot-Watt University, Edinburgh EH14 4AS, United Kingdom}

\begin{abstract}
Periodically-driven quantum systems can exhibit topologically non-trivial behaviour, even when their quasi-energy bands have zero Chern numbers. Much work has been conducted on non-interacting quantum-mechanical models where this kind of behaviour is present. However, the inclusion of interactions in out-of-equilibrium quantum systems can prove to be quite challenging. On the other hand, the classical counterpart of hard-core interactions can be simulated efficiently via constrained random walks. The non-interacting model proposed by Rudner {\it et al.} [Phys. Rev. X {\bf 3}, 031005 (2013)], has a special point for which the system is equivalent to a classical random walk. We consider the classical counterpart of this model, which is exact at a special point even when hard-core interactions are present, and show how these quantitatively affect the edge currents in a strip geometry. We find that the interacting classical system is well described by a mean-field theory. Using this we simulate the dynamics of the classical system, which show that the interactions play the role of Markovian, or time dependent disorder. By comparing the evolution of classical and quantum edge currents in small lattices, we find regimes where the classical limit considered gives good insight into the quantum problem.
\end{abstract}

\maketitle

\section{Introduction}

The study of topologically protected properties of quantum systems has been an area of increasing interest in recent years \cite{Kitagawa2010,Gold2012,Lababidi2014,Atala2014,Reichl2014}. This started from the discovery of the quantum Hall effect \cite{Klitzing1980} and has continued to the more recent discovery of topological insulators \cite{Hasan2010,Murakami2011}. As an explanation of the quantum Hall effect, Halperin put forward the idea of current-carrying edge states \cite{Halperin1982}. Non-equilibrium states localised along the edge have been recently theoretically developed using periodically driven quantum systems \cite{Rudner2013,Gold2014,Asboth2014}. 

In the type of model we consider the lattice can support chiral edge modes which are robust against disorder. These edge modes can arise in band insulating and superconducting systems \cite{Kitagawa2010,Sacramento2015}. The Chern number of the energy bands can be used in static systems to predict the number of edge modes present \cite{Schnyder2008}. However, in a driven system the Chern number does not always give enough information to characterise the number of edge modes \cite{Rudner2013}, and a new topological invariant is needed. In a periodically driven non-interacting system it has been shown that such an invariant can be constructed and the existence of edge modes in such a system predicted \cite{Rudner2013}, even for a system with all Chern numbers equal to zero.

The introduction of interactions into periodically driven systems can be difficult. However, some properties have been considered, for example the calculation of effective Hamiltonians for 1D lattices with high driving frequency \cite{Itin2015}, the classification of their topological phases \cite{Else2016}, or the emergence of a many-body localized phase in the prescence of disorder \cite{Ponte2015a,Ponte2015b}. In this work we focus on the effect of interactions on the edge current of a periodically driven topological insulator. The stability of edge modes in a photonic Floquet topological insulator with nonlinearities (interactions) present has been considered in ref \cite{Lumer2016}. Single edge modes were found to be unstable upon the introduction of these interactions. Here we consider the collective behaviour of the current along the edge of the system in the presence of interactions.

In this paper we investigate the classical limit of a periodically driven model possessing robust chiral edge states \cite{Rudner2013}. For the classical limit of the system, the quantum dynamics are substituted by a classical tunnelling probability, where any quantum phases are neglected. From this we hope to gain insight into the properties of the quantum model with interactions. Strictly speaking this is a comparison between discrete time random walks \cite{Angstmann2015} and quantum walks \cite{Asboth2015,Genske2013,Kitagawa2012}. Classical and quantum random walks can have starkly different behaviour \cite{Childs2002,Boetteher2015}. It has also been shown that discrete time quantum walks can have rich topological phases \cite{Kitagawa2010,Asboth2013}. There has been much work on the quantum to classical random walk transition \cite{Kendon2003,Brun2003,Oliveira2006,Kosik2006,Whitfield2010,Preiss2015}, including an experimental implementation using photons \cite{Broome2010}. Quantum walks with small amounts of decoherence give classical random walk results \cite{Brun2003,Oliveira2006,Kosik2006}. Therefore our classical limit can be seen to give a tenuous insight into broad general properties of edge modes in the interacting quantum model, and a strong indication of the edge current properties of driven interacting quantum systems where decoherence is present.

To take the classical limit we simply neglect any phases of the system. This is equivalent to taking random phase shifts after each step of the model, which results in a transition from the quantum to the classical walk \cite{Kosik2006}. We note that the classical limit of the model in Ref.~\cite{Rudner2013} is equivalent to its quantum counterpart at a special point, where the tunnelling probability after one quarter of a period is unity, even in the presence of hard-core interactions.

In Sec.~\ref{sec:Model} and Sec.~\ref{sec:Classical} we introduce the quantum and classical models we will consider. In Sec.~\ref{sec:DefinitionCurrent} we define an edge current observable. We then move on to discuss a theoretical estimation of the edge current for the classical limit in Sec.~\ref{sec:ApprModel}. This is followed by a discussion of the simulations conducted in the classical limit. In Sec.~\ref{sec:InterDisorder} we investigate the effect of interactions on the edge current of our system, and its relation to time-dependent disorder. We then compare the dynamics of the edge current for the quantum and classical model for small lattices in Sec.~\ref{sec:QuantClass}. In the last part of this work, Sec.~\ref{sec:ComparCurrent}, we compare the different current observables discussed in Sec.~\ref{sec:DefinitionCurrent} for the classical and quantum models.

\section{Quantum mechanical model}
\label{sec:Model}
\begin{figure}[t]
\begin{center}
\includegraphics[width=0.45\textwidth]{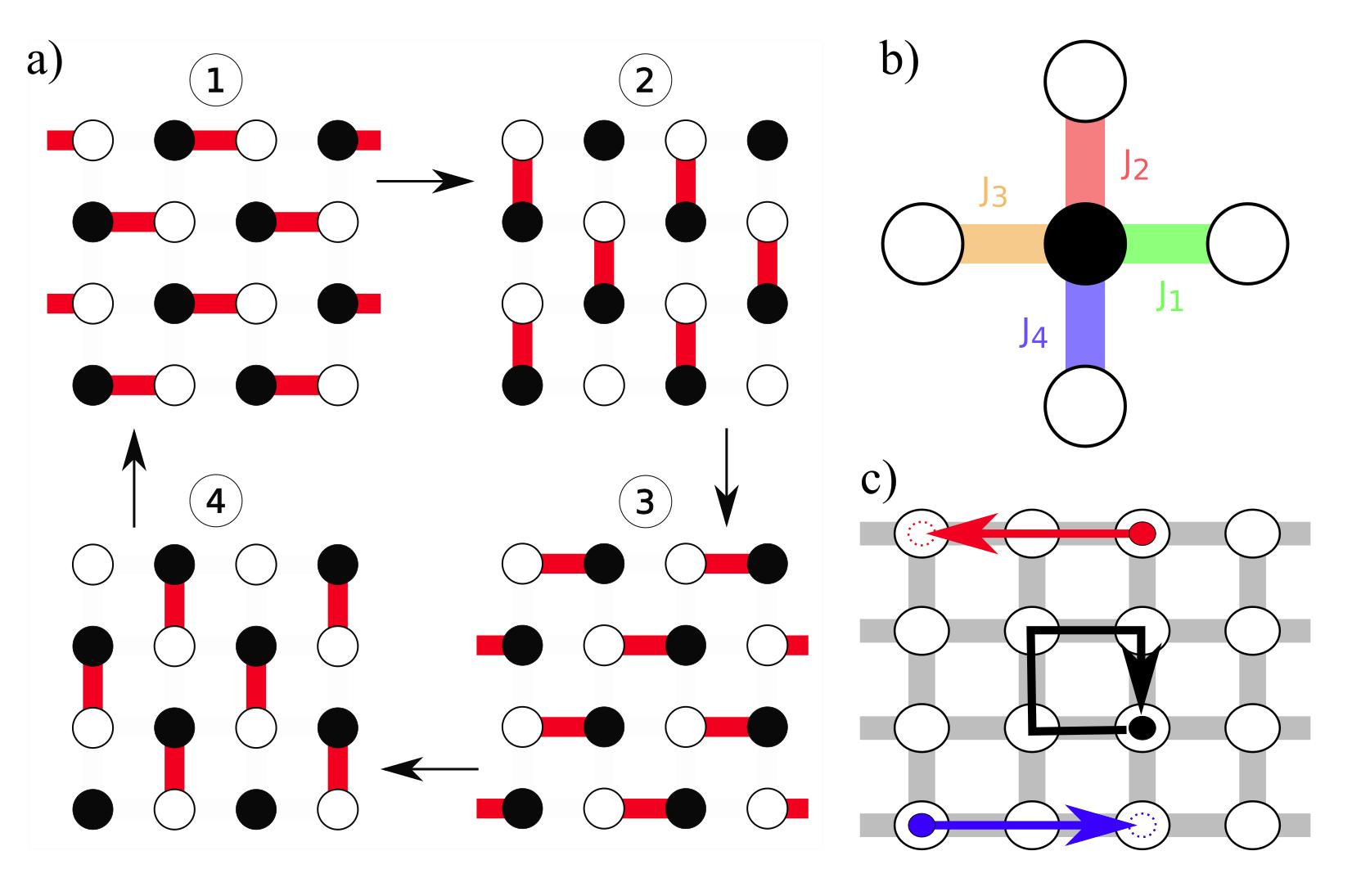}
\end{center}
\caption{Illustration of the model. a) Steps for the periodic driving of this model. At each step the sites that are coupled are joined by red lines. Sublattice A(B) is drawn as a white(black) site. The lattice is periodic in the horizontal direction but has horizontal edges. b) The order of couplings from a single B site to the adjacent A sites. c) Illustration of edge (indicated by red and blue arrows) and bulk paths (black arrow) for unit tunnelling probability.}
\label{fig:Model}
\end{figure}

Here, we present a two-dimensional tight-binding model on a square lattice that has non trivial topological properties, and is almost identical to the simple model proposed by Rudner {\it et al.} \cite{Rudner2013}. This, in spite of being just a toy model in condensed matter physics, has been recently realised experimentally, in the single-particle case, using light in laser-written waveguides \cite{Mukherjee2016,Maczewsky2016}. The non-interacting dynamics of this system is described by the following time-dependent Hamiltonian
\begin{equation}
H(t)=\sum_{\langle \mathbf{i},\mathbf{j} \rangle}J_{\mathbf{i},\mathbf{j}}(t)\ket{\mathbf{i}}\bra{\mathbf{j}} + \mathrm{H.c.},
\label{eqn:Hamiltonian}
\end{equation}
where $\mathbf{i}=(i_x,i_y)\in \mathbb{Z}^2$ labels the lattice sites, and $\langle \mathbf{i},\mathbf{j}\rangle$ restricts the sum over nearest-neighbours. The tunnelling rates $J_{\mathbf{i},\mathbf{j}}(t)$ are time-periodic with period $T$, i.e. $J_{\mathbf{i},\mathbf{j}}(t+T)=J_{\mathbf{i},\mathbf{j}}(t)$. The tunnelling rates are activated sequentially in $4$ steps as follows (see Fig.~\ref{fig:Model}a):
\begin{equation}
\begin{aligned}
J_1 \colon & J_{\mathbf{i},\mathbf{j}\pm \hat{e}_y} =0 \mbox{, } 2J_{\mathbf{i}+\hat{e}_x,\mathbf{j}}= [1-(-1)^{i_x+i_y}] J \\ & \mbox{ for } mT < t \leq mT + T/4 \\
J_2 \colon & J_{\mathbf{i}\pm \hat{e}_x,\mathbf{j}}=0 \mbox{, } 2J_{\mathbf{i},\mathbf{j}+\hat{e}_y}= [1+(-1)^{i_x + i_y}] J  \\ & \mbox{ for }  mT + T/4 < t \leq mT + T/2 \\
J_3 \colon & J_{\mathbf{i},\mathbf{j}\pm \hat{e}_y} =0  \mbox{, } 2J_{\mathbf{i}+\hat{e}_x,\mathbf{j}}= [1+(-1)^{i_x + i_y}] J \\ & \mbox{ for } mT + T/2 < t \leq mT + 3T/4 \\
J_4 \colon & J_{\mathbf{i}\pm \hat{e}_x,\mathbf{j}}=0 \mbox{, } 4J_{\mathbf{i},\mathbf{j}+\hat{e}_y}= ( [1-(-1)^{i_x + i_y}] J \\ & \mbox{ for } mT + 3T/4 < t \leq mT + T\\
\end{aligned}
\label{Eq2}
\end{equation}
Above, $m$ is any integer, $J$ is a real constant and we have defined $\hat{e}_x=(1,0)$ and $\hat{e}_y=(0,1)$.

\begin{figure}[h]
\centering
\includegraphics[width=0.49\textwidth]{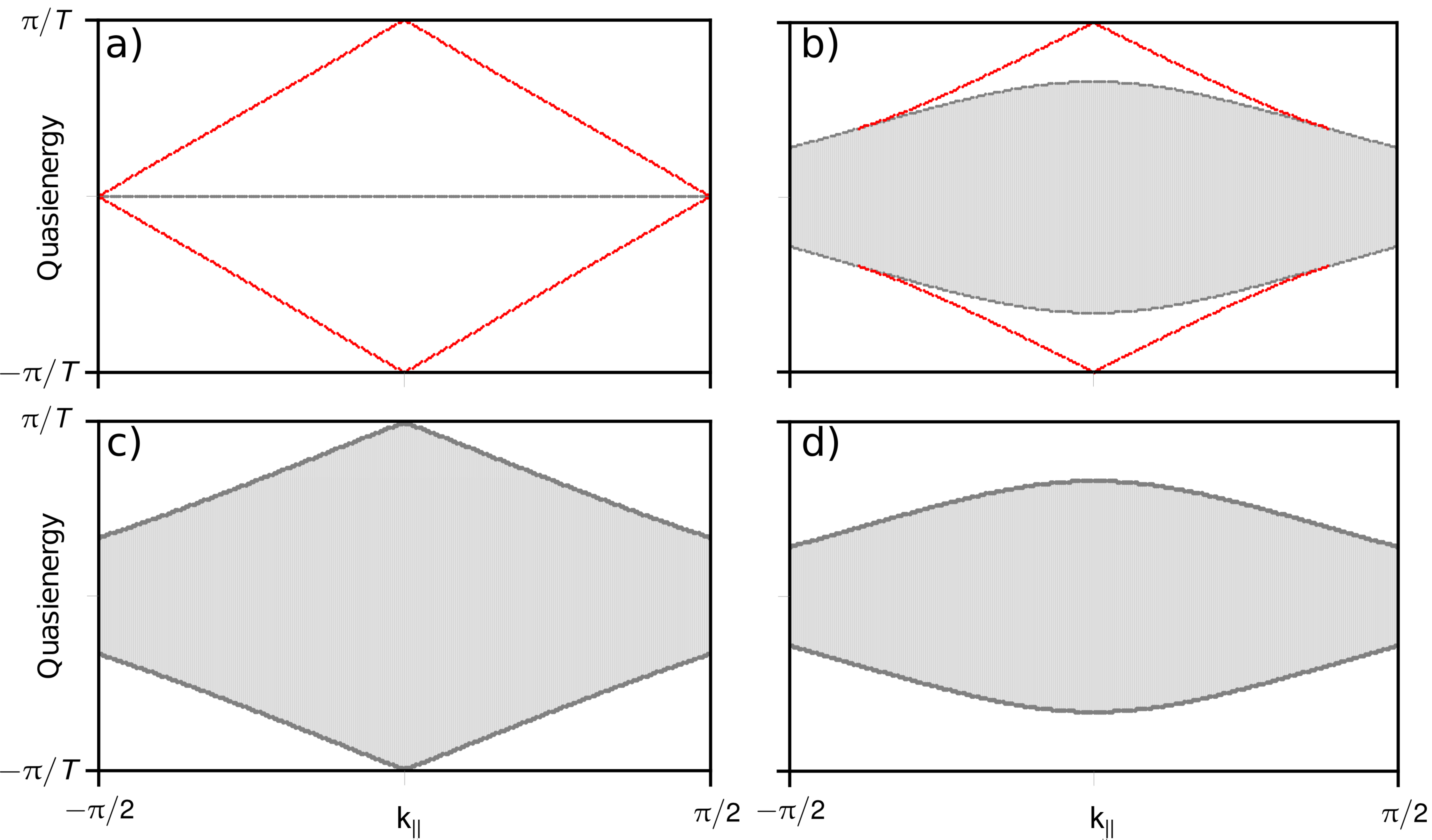}
\caption{Spectra of the quantum model. a) $J = \frac{\pi}{2}$ ($P_t\sim 100\%$) b) $J = 1.05$ ($P_t\sim75\%$) c) $J = \frac{\pi}{4}$ ($P_t\sim 50\%$) d) $J = 0.52$ ($P_t\sim 25\%$).}
\label{fig:Spectra}
\end{figure}

The spectrum of this model confirms the presence of edge modes \cite{Rudner2013}, see Fig.~\ref{fig:Spectra}. As expected for the special case of $J = \pi / 2$  (with a cumulative phase of $4\times \pi/2 = 2 \pi$ in one cycle) we get bulk modes in closed trajectories and ``perfect" (i.e. purely linearly-dispersed) edge modes \cite{Rudner2013,Murakami2011}. As we decrease the tunnelling rates the edge mode dispersion attains some curvature, and they are completely unobserved in the topologically trivial case \cite{Reichl2014}, Fig.~\ref{fig:Spectra}d.

\section{Classical model}
\label{sec:Classical}

As pointed out in the previous section, the quantum model has a special point, $J= \pi/2$ (Fig.~\ref{fig:Spectra}a), at which the tunnelling probability after each of the tunnelling sequences between two connected sites is unity. Therefore, if we look at a single particle with initial wavefunction $\ket{\psi(0)}=\ket{\mathbf{i}_0}$ at discrete times $t_m = mT/4$, its dynamics is identical to a classical random walk with four distinct steps defined at times $t_m$, as in Eq.~(\ref{Eq2}). To define the classical counterpart of the quantum model, for general $J$, we ignore the quantum-mechanical nature of the tunnelling and substitute the quantum dynamics by a classical tunnelling probability $P_t=P_t(J)$ at each discrete time $t_m$. For a single particle, this classical model neglects quantum phases and only gives the correct time-dependent probabilities $P(\mathbf{i},\mathbf{i}_0;t_m)\equiv P_m(\mathbf{i},\mathbf{i}_0)$ that a particle with initial state $\ket{\mathbf{i}_0}$ is found at $\ket{\mathbf{i}}$ after $m$ time steps for $P_t=1$ (or $P_t=0$). However, for tunnelling probabilities sufficiently close to $1$ or $0$ and for sufficiently short times of order one period, the classical probabilities are  a good approximation, and we have
\begin{equation}
P_m(\mathbf{i})\approx \left|\langle \mathbf{i}|\psi_{\mathbf{i}_0}(t_m)\rangle\right|^2.
\end{equation}    

We can formally obtain a form for the relationship $P_t = P_t(J)$, by exploiting the fact that the sites are only coupled in pairs at any one time step of the driving, see Fig.~\ref{fig:Model}a. The quantum Hamiltonian for a single pair can be denoted by
\begin{equation}
H = \begin{pmatrix}
0 & J \\ J & 0
\end{pmatrix}.
\label{eqn:PairHamil}
\end{equation}  
Above, we have defined the wavefunction for the pair to be given as
\begin{equation}
\psi(t) = \begin{pmatrix}
c_1(t) \\ c_2(t)
\end{pmatrix},
\label{eqn:PaitWavefunc}
\end{equation}
where $c_{1(2)}(t)$ is the occupation of site $1(2)$ at time $t$. We begin with a state of the form $\psi(0) = (1,0)$. We then solve the Schr\"{o}dinger equation after one time step $t_i$, and get a probability density of
\begin{equation}
\begin{pmatrix}
\mid c_1(t) \mid^2 \\ \mid c_2(t) \mid^2
\end{pmatrix} = \begin{pmatrix}
\cos^2(J t_i) \\ \sin^2 (J t_i)
\end{pmatrix}.
\label{eqn:FinalWavefunc}
\end{equation}
The relationship between the classical probability and quantum tunnelling rates becomes
\begin{equation}
P_t  = \sin^2 (J t_i)
\label{eqn:ProbRelation}
\end{equation}
For example, to obtain complete transfer in time $t_i$ we require that $J t_i = \pi / 2$. During this work the time for each step is set to $t_i = 1$.

To illustrate the comparison between the classical and quantum probabilities, we calculate the probability of a walker either staying or coming back to its initial location $\mathbf{i}_0$ after a complete period $T$, given an initial state $\ket{\psi(0)}=\ket{\mathbf{i}_0}$. The quantum dynamics is given by $\ket{\psi(t_1)}=(A\ket{\mathbf{i}_0}+B\ket{\mathbf{i}_1})/\sqrt{(|A|^2+|B|^2)}$, where $\mathbf{i}_1$ is the only site connected to $\mathbf{i}_0$ in time step $t_1$. We define $A=|A|\exp(i\phi_A)$, $B=|B|\exp(i\phi_B)$ with $\phi_A$($\phi_B$) being the phase gained on the walker remaining(tunnelling) after one time step. The probability of the walker being in its initial state after one period is given by
\begin{equation}
\begin{aligned}
\left|\langle \mathbf{i}_0|\psi_{\mathbf{i}_0}(t_4)\rangle\right|^2&=P_t^4+(1-P_t)^4\\&+2P_t^2(1-P_t)^2\cos(\phi_S-\phi_L),
\label{eqn:WalkerProb}
\end{aligned}
\end{equation}
where $\phi_S = 4\phi_A$ ($\phi_L = 4\phi_B$) is the accumulated phase for the walker remaining stationary (returning) in one period. From Eq.~(\ref{eqn:WalkerProb}), we see that phase differences of $\pi/2$ also recover the classical result.

We discuss now the interacting many-particle case. We consider $N$ particles with identical tunnelling probabilities $P_t$ at each time step. Moreover, we consider the particles to be indistinguishable, in the sense that we do not label each individual particle, but only care about whether a site is empty or occupied. To simulate a hard-core interaction, our initial states do not contain two or more particles on the same lattice site, nor allow for multi-particle occupancy of any site at any time. The particles live on a square lattice with a strip geometry, i.e. periodic boundary conditions in the $x$-coordinate and open boundary conditions in the $y$-coordinate. We choose the lattice to have the same dimension $L$ in both directions, and the filling factor is given by $\nu = N/L^2$. To see how the hard-core constraint operates in this system, it suffices to consider two particles on neighbouring sites at a discrete time when these two sites are connected. If we label such a two-body state at time $t_0$ as $\ket{\mathbf{i},\mathbf{j}}$, then the state at time $t_0+T/4$ does not change, regardless of $P_t$, due to the indistinguishably of the particles. In any other case, i.e. when one site is occupied and its connected site is empty, the particle undergoes its single-particle random walk with tunnelling probability $P_t$.

\section{Edge Current Observables}
\label{sec:DefinitionCurrent}

We first need an appropriate definition for the current $\mathcal{I}$ along the edge. This is not straightforward, as we have a flow of particles in and out of the edge, resulting in the particle number not being conserved along the edge. To measure the flow we consider the possibility of having a (non-destructive) detector between each pair of lattice sites along the edge. This is inspired by techniques used in the ultracold atom community of projecting states into isolated wells \citep{Atala2014,Hugel2014,Kessler2014}. A simple definition of current could then be given by counting the number of tunnelling events moving left $N_{L}$ and right $N_{R}
$ along the edge. After a number, $m$, of periods, $T$, we can define $\mathcal{I}$ simply as
\begin{equation}
\mathcal{I} = \frac{N_{L}-N_{R}}{LmT},
\label{eqn:Current}
\end{equation}
with $L$ denoting the length of the lattice along the edge.

Naturally, for averaging over random initial states the current, as given by Eq. (\ref{eqn:Current}), will be zero. This is a consequence of having the lattice randomly and uniformly filled.

We need to define a different measure of the edge current, which gives us information about the dynamics of the system if we choose random initial states. The measure that we consider is the component of the current in one direction. First we take advantage of the symmetry of the model to simplify notation by considering only the current on one of the edges (top edge) of the strip geometry. We then measure on this edge the flow in the leftward direction, which is the direction that the main edge path moves. We will refer to this as the directional edge current $\mathcal{J}$, and a simple definition is given by
\begin{equation}
\mathcal{J} = \frac{N_{L}}{LmT}.
\label{eqn:CurrentEdge}
\end{equation} 
Note that the directional current is always positive by construction because we are measuring the flow in only one direction. If the driving of Fig.~\ref{fig:Model}a was reversed, then our directional edge current will not change sign and would continue to measure the current in the same direction.

We consider initial, uniformly filled random states with filling $\nu$ of indistinguishable particles, which in the classical case correspond to not labelling. We can write a more general definition of Eqs.~(\ref{eqn:Current}) and~(\ref{eqn:CurrentEdge}) in terms of the occupation of sites at specific times. To write the $\mathcal{I}$ and $\mathcal{J}$ in this form we can take advantage of the fact that movement along the edge can only occur during the first and third time steps, Fig.~\ref{fig:Model}a. In each of these time steps the system breaks up into a set of pairs and we can get the directional edge current from the occupation of only the left hand sites before $t_{u-1}$ and after the time step $t_u$, where we have denoted the step as $u$. We will notate the occupation of the left site of the pair as $n_{\mathrm{Left}}$. For the classical simulations this can have the value of $1$ or $0$, i.e. the site is occupied or empty. Note that this will not be the case in Sec.~\ref{sec:InterDisorder}, where we consider mean-field theory, nor in Sec.~\ref{sec:QuantClass}, where we consider the quantum case. For each pair and for each time step we can then measure the edge current $\mathcal{I}$ by
\begin{equation}
\mathcal{I}\left(t_u\right) = \frac{n_{\mathrm{Left}}(t_u) - n_{\mathrm{Left}}(t_{u-1})}{LmT},
\label{eqn:GeneralCurrent}
\end{equation}
with $m$ denoting the total number of time periods the system will be evolved through and $T$ the period of the driving. The directional edge current $\mathcal{J}$ is given by,
\begin{equation}
\begin{aligned}
\mathcal{J}\left(t_u\right) = & \frac{\left(n_{\mathrm{Left}}(t_u) - n_{\mathrm{Left}}(t_{u-1})\right)}{LmT} \\ & \times \frac{1 + {\mathrm{sgn}}\left( n_{\mathrm{Left}}(t_u) - n_{\mathrm{Left}}(t_{u-1}) \right)}{2}.
\end{aligned}
\label{eqn:GeneralCurrentEdge}
\end{equation}
Above, $\mathrm{sgn}(x)$ denotes the sign of $x$. Hence, we have the $\mathcal{I}$ and $\mathcal{J}$ in terms of measurable quantities in both classical and quantum models. The total current in one time period is the sum of Eq.~(\ref{eqn:GeneralCurrent}) over all pairs of coupled sites on the edge during driving steps one and three. We have defined the edge current above such that leftward motion results in a positive current.

\section{Theoretical estimation of edge currents in the classical limit}
\label{sec:ApprModel}

Before discussing the simulations of the classical model, we present an approximate theory that describes quantitatively the dynamics for all filling factors. In particular, we are interested in the directional currents flowing at the two edges of the strip geometry which are, on average, equal in magnitude and flow in opposite directions. We consider random initial states with no multi-particle initial occupancies at any lattice site, and calculate the edge currents, averaged over all possible initial states and in time.

\begin{figure}[h]
\centering
\includegraphics[width=0.49\textwidth]{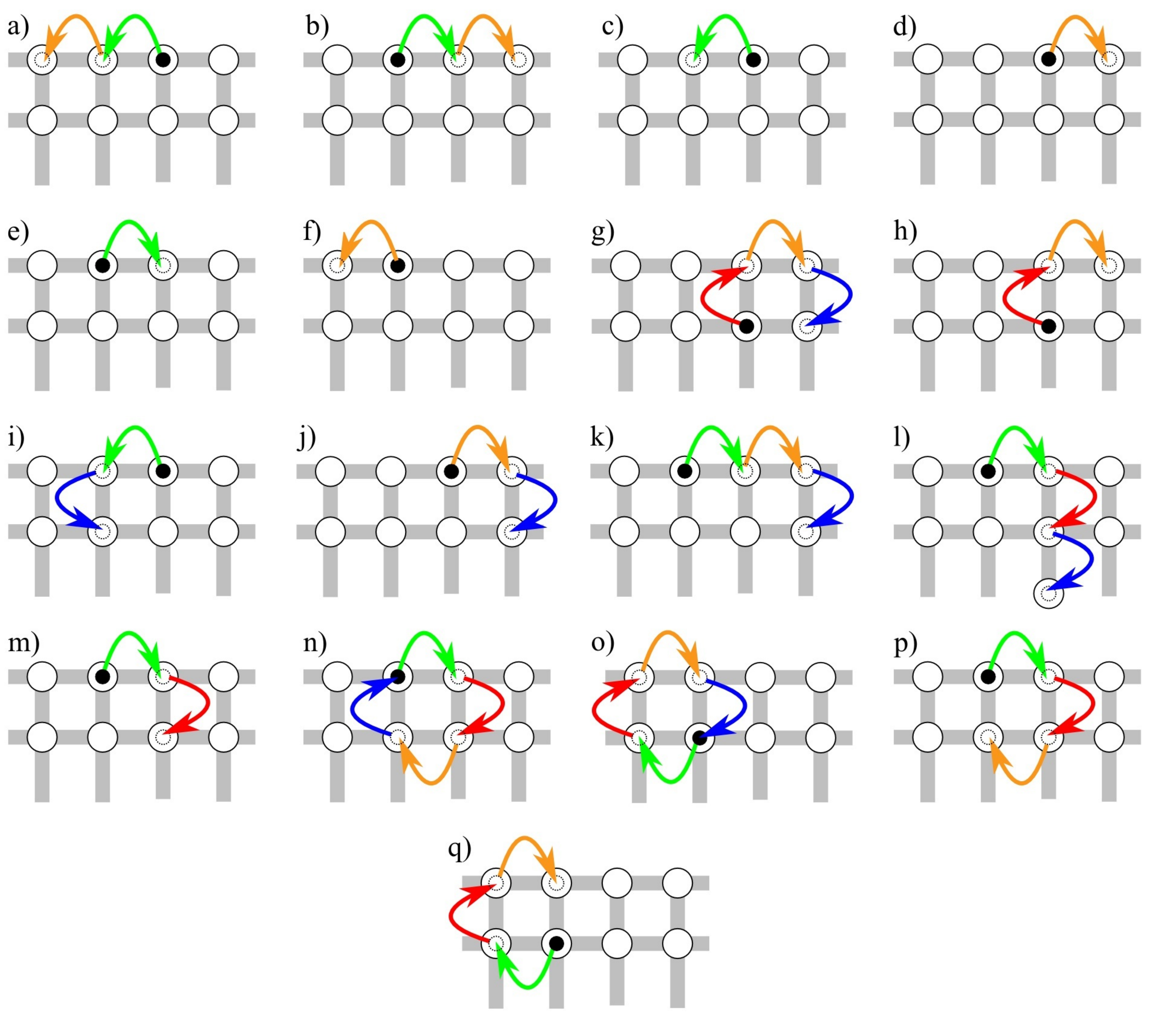}
\caption{All possible one-body paths that contribute to the current, Eq.~(\ref{eqn:GeneralCurrent}). The arrows are colour coded to the step they occur in our considered driving cycle as in Fig.~\ref{fig:Model}b.}
\label{fig:SinglePart}
\end{figure}

To calculate the edge current, we need to consider all the possible paths for a particle to move in the $x$ direction along the edge in one period of our driving protocol. Each of the 17 paths that contribute are shown in Fig.~\ref{fig:SinglePart}, with a mixture of edge and bulk paths contributing to the current. We define an edge path as a particle movement that is entirely confined to the edge and a bulk path as a particle movement in the bulk that includes a component of movement along the edge. To find the form of the overall current we need to calculate each of the 17 paths respective probabilities of occurring, including the tunnelling probability and the hard-core interacting constraint. Below, we begin with discussing how we define the probabilities of occurrence of a path and then move on to look in detail at the non-interacting and strongly interacting cases.

\subsection{Probability of Current Contributing Walks}
\label{sec:Contib}

We describe now how the probability for each path to take place,  $\mathit{p}_i$, can be calculated. We consider the driving protocol with a probability of tunnelling in each time step and take into account the occupancy of the site being coupled to, over all four time steps that constitute a single period. It is possible to do this by separating each path into its four sequential steps. In the calculations of these probabilities and the resulting currents (see appendix~\ref{sec:AppPaths}) we assume the particles are randomly and uniformly distributed and that no correlations of their positions build up over large time scales. The assumption of no correlations built up over time is only true on average over the many random initial states, with any correlations being washed out. These are reasonable assumptions for a classical particle, where we have no superposition or interference. These assumptions would not necessarily be valid for a quantum particle. However, they give a good estimation of the resulting edge currents in the classical model, as we will show when discussing the simulations conducted in Sec.~\ref{sec:Sim}.

First, we consider the main edge path, Fig.~\ref{fig:SinglePart}a, which for each step has the probability, for non-interacting (NI) and strongly interacting (SI) particles,
\begin{equation}
\begin{aligned}
& \mbox{NI} &  \mbox{SI} \\
\mbox{1: } & P_t  &  P_t \left(1-\nu\right) \\
\mbox{2: } & 1  &  1  \\
\mbox{3: } & P_t  &  P_t \left(1-\nu\right) \\
\mbox{4: } & 1  &  1 \\
\end{aligned}
\end{equation}
The particle tunnels in steps 1 and 3. This occurs with a probability $P_t$ and for the strongly interacting case the site the particle is going to tunnel to needs to be empty, giving a $\left(1-\nu\right)$ term. In steps 2 and 4 the particle is not coupled to any sites, as can be seen from Fig.~\ref{fig:Model}a, resulting in a probability of unity for the particle staying in the site. Combining these steps we get for non-interacting particles a path probability of $\mathit{p}^{NI}_a = \nu P_t^2$ and for strongly interacting particles $\mathit{p}^{SI}_a = \nu P_t^2 \left(1-\nu\right)^2$. The extra $\nu$ term in each of these expressions is a result of requiring the initial starting site to be occupied.

The path given in Fig.~\ref{fig:SinglePart}b is more complex. At each step this path has the probabilities of
\begin{equation}
\begin{aligned}
& \mbox{NI} &  \mbox{SI} \\
\mbox{1: } & P_t  &  P_t \left(1-\nu\right) \\
\mbox{2: } & \left(1-P_t\right)  &  \left(\nu + \left(1-\nu\right)\left(1-P_t\right)\right)  \\
\mbox{3: } & P_t  &  P_t \left(1-\nu\right) \\
\mbox{4: } & \left(1-P_t\right)  &  \left(\nu + \left(1-\nu\right)\left(1-P_t\right)\right) \\
\end{aligned}
\end{equation}
For steps 1 and 3 we have the same as for the previous path, but now for steps 2 and 4 the site containing the particle is coupled to another but the particle does not tunnel for the required path. For the non-interacting case this can only occur with probability $\left(1-P_t\right)$. However, for interacting particles there are two ways the particle might not tunnel: the site can be occupied ($\nu$) or it is not occupied and the particle does not tunnel ($\left(1-\nu\right)\left(1-P_t\right)$). For non-interacting particles we then get a path probability of $\mathit{p}^{NI}_b = \nu P_t^2\left(1-P_t\right)^2$ and for strongly interacting particles $\mathit{p}^{SI}_b = \nu P_t^2 \left(1-\nu\right)^2\left(\nu + \left(1-\nu\right)\left(1-P_t\right)\right)^2$.

With the building blocks described above we have all we need to define the probability for each of the 17 possible paths to occur. We also need the total number of times this path is possible in the lattice. For all 17 paths this has the same value, $L/2$, coming from the fact that we have two different sublattices, referred to as A and B in Fig.~\ref{fig:Model}. Each of the paths is unique to one sublattice, and in the lattice there are $L/2$ sites of each sublattice in each row. The second term to consider is the total current contributed by that path. This is given by the total tunnelling events along the x-direction while on the edge. For path a) this is +2 and for path b) this is -2.

\subsection{Non-Interacting Particles}
\label{sec:modelNonInter}

For the non-interacting case we can calculate the contributing current for each path from Fig.~\ref{fig:SinglePart}, given in Eq.~(\ref{eqn:PathsNonInter}). Summing the contribution from all paths, we obtain that there should be zero current along the edge, $\mathcal{I} = 0$. This may be surprising at first but we note that the edge properties of the system strongly depend on the initial state. Since our initial state is completely random, the current must be zero. This is the argument we used in the Sec. \ref{sec:DefinitionCurrent} for the requirement of the definition of another observable, the directional edge current $\mathcal{J}$, in order to be able to monitor the system's dynamics for random initial states.

Using the definition of $\mathcal{J}$ in Eq.~(\ref{eqn:CurrentEdge}), we can find an expression for the expected directional edge current in the non-interacting regime,
\begin{equation}
\mathcal{J}_{NI} = \nu P_{t}. 
\label{eqn:NonInter}
\end{equation}
This is obtained by the sum of all the currents that contribute to the directional edge current from Eq.~(\ref{eqn:PathsNonInter}) in Appendix~\ref{sec:AppPaths}. As is clear from Eq.~(\ref{eqn:NonInter}), the directional edge current $\mathcal{J}_{NI}$ grows linearly with the filling factor $\nu$, as one expects for a non-interacting system. The situation is however different for strong interactions.

\subsection{Strongly Interacting Particles}
\label{sec:modelInter}

Summing all the contributing paths for strongly interacting particles, we obtain the directional edge current in the strongly interacting regime
\begin{equation}
\mathcal{J}_{SI} = \nu P_{t} \left(1-\nu\right). 
\label{eqn:Inter}
\end{equation}
Again this is obtained by the sum of all the currents that contribute to the directional edge current from Eq.~(\ref{eqn:PathsInter}) in Appendix~\ref{sec:AppPaths}. As expected the directional edge current $\mathcal{J}_{SI}$ does not grow linearly with the filling factor $\nu$. Instead we get a maximum directional current at $\nu = 0.5$ with it going to zero at $\nu = 1$. The directional edge current goes to zero due to a jamming of the system which is the result of the hard-core constraint. This is analogous to the jamming ocurring in traffic flow models \cite{Chowdhury2000}.

\subsection{Mean Field Directional Edge Current}

There is a simple map of the probability between the non-interacting and strongly interacting systems directional edge currents, Eq.~(\ref{eqn:NonInter}) and Eq.~(\ref{eqn:Inter}) respectively, that is, between the cases considered in Sec.~\ref{sec:modelNonInter} and Sec.~\ref{sec:modelInter}. This can be expressed as a transform of the tunnelling probability in the non-interacting case,
\begin{equation}
P_t^{NI} \rightarrow P_t \left(1-\nu\right).
\label{eqn:Map}
\end{equation}
Hence, in the strongly interacting regime the mean-field description gives a modification of the tunnelling probability which is captured by a $(1-\nu)$ ``excluded volume" term. This allows for the strongly interacting system to be simulated with a non-interacting one, i.e. a classical random walk.

We can then run single-particle simulations, but with a rescaled tunnelling probability, Eq.~(\ref{eqn:Map}). This results in a single particle, interacting with a mean-field sea of randomly and uniformly distributed hardcore particles, with a density $\nu$. We will exploit this in Sec.~\ref{sec:InterDisorder}, to look into the time dynamics of a single particle in the strongly interacting regime.

For the mean-field case it is assumed that all particles are randomly and uniformly distributed. We can assume this at all times for the classical case, as there are no long time correlations that build up. This would not be the case in the quantum problem. Where the build up of phases results in superpositions and interference that get more complex for longer times.

\section{Simulation of the Directional Edge Current}
\label{sec:Sim}

We have run simulations in an $L\times L$ lattice on a strip geometry, with $N$ particles from 1 to $L^2-1$. We fix the size of the system to $L=100$, which is sufficiently large to avoid finite size effects while remaining manageable from a computational time point of view. For each value of the tunnelling probability $P_t$, we run $s=5000$ simulations with random starting configurations for the $N$ particles with at most one particle per site, for a number $m= 100$ of periods. The steps of particles in either direction along the two edges is counted. This allows us to count the full flow as defined by Eq.~(\ref{eqn:GeneralCurrent}) and then measure the directional edge current as defined in Eq.~(\ref{eqn:GeneralCurrentEdge}). All currents are averaged over the simulated initial configurations.

\begin{figure}[h]
\centering
\includegraphics[width=0.49\textwidth]{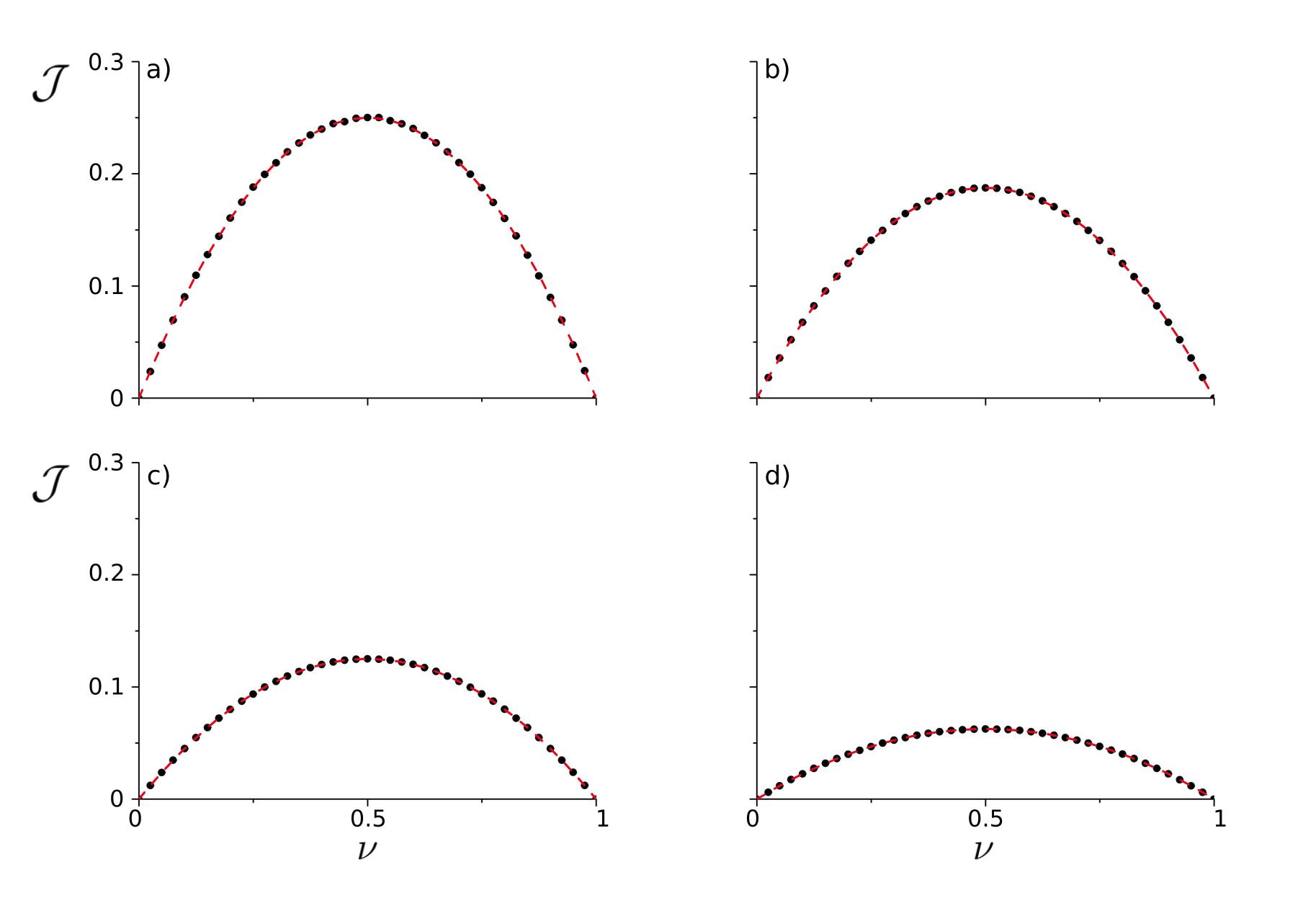}
\caption{Directional edge currents in the strongly interacting case. (Black) points are from the numerical simulations and the dashed (red) line is the theoretical estimation from Eq.~(\ref{eqn:Inter}). a)--d) correspond to different values of $P_t$, a) $P_t = 100\%$, b) $P_t = 75\%$, c) $P_t = 50\%$, d) $P_t = 25\%$.}
\label{fig:Interacting}
\end{figure}

We first compare the simulations and theoretical estimation for strongly interacting particles, seen in Fig.~\ref{fig:Interacting}. There is perfect agreement between our estimation and the measured directional current from the simulations. The resulting directional edge currents are symmetric around $\nu = 1/2$, as expected from the particle-hole symmetry of the problem. This is to say that with our driving cycle a single hole on an edge full of particles would flow in the same way as a single particle on the edge.

A comparison can be drawn between the form of our results for the strongly interacting regime and the flows observed in the Nagel---Schreckenberg cellular automata model of traffic \cite{Schadschneider1999,Chowdhury2000,Schadschneider2002}. The Nagel---Schreckenberg model starts with a random configuration of cars of density $\rho$ along a road, that is split into discrete sections. In each discrete section one car is allowed to be present. There are 4 steps in the evolution of the model: 1) vehicles accelerate, 2) vehicles decelerate due to the other cars, 3) a random deceleration with a certain probability and 4) each vehicle moves forward due to its current velocity \cite{Nagel1992}. The update procedure is usually conducted for each vehicle in parallel to account for the rich dynamics of the system. However, a ``mean-field" result can be calculated for the flow in the case of updating one vehicle at a time to be \cite{Schadschneider2002,Nagel1992}
\begin{equation}
\mathcal{J}_{NaSch} = \rho \left(1 - \rho \right).
\label{eqn:Traffic}
\end{equation}
This follows the exact relationship of our interacting model as seen in Eq.~(\ref{eqn:Inter}). This is unsurprising as the problem of traffic flow is intrinsically linked to the properties of transport of classical strongly interacting particles. They share the important properties of being classical (no superpositions or interference of vehicles), and having a maximum occupancy per unit cell of one. There has been a lot of work to understand and extend traffic flow cellular automata in more detail \cite{Chowdhury2000,Schadschneider2002,Maerivoet2005}, including particle hopping models \cite{Nagel1996}.

We can draw a comparison between our results and the quantum model spectra from Fig.~\ref{fig:Spectra}. In the spectra it was observed that the edge modes were lost into the bulk when the coupling between sites was reduced. In the classical limit a lowering of the tunnelling probability does give a linear decrease in the observed directional edge currents. This would not be the case in the quantum problem, as we will see in Sec.~\ref{sec:QuantClass}, where the build up of phases results in superpositions and interference which gives rise to more complex dynamics for longer times.

\section{Interaction effects on edge current and Markovian disorder}
\label{sec:InterDisorder}

In this section we consider the classical analog of monitoring the time dynamics of a single distinguishable particle being initially prepared in an edge or bulk mode, with strong interactions with the rest of the particles, whose dynamics are ``traced" away. We define an edge mode as a single particle being launched on the edge of the system in a site of type A. This is reasonable since it results in flow along the edge at the special point of unit tunnelling. A bulk mode is defined as a single particle being launched in any other site. The dynamics of this single particle is tracked. It is possible to look at the dynamics of this system due to the mean-field description obtained in Sec.~\ref{sec:modelInter}, Eq.~(\ref{eqn:Map}). For simplicity, and without loss of generality, we will consider $P_t=1$, and vary the filling factor $\nu$ of the system since, as we have seen, mean-field theory appears to be exact for the directional edge current, and with the initial states we are considering. The mean-field theory  corresponds to the replacement $P_t\to P_t(1-\nu)$. We will also establish the relationship between interaction effects and Markovian, or time-dependent disorder, and highlight some differences with non-Markovian, or static (quenched) disorder. 

\begin{figure}[h]
\centering
\includegraphics[width=0.4\textwidth]{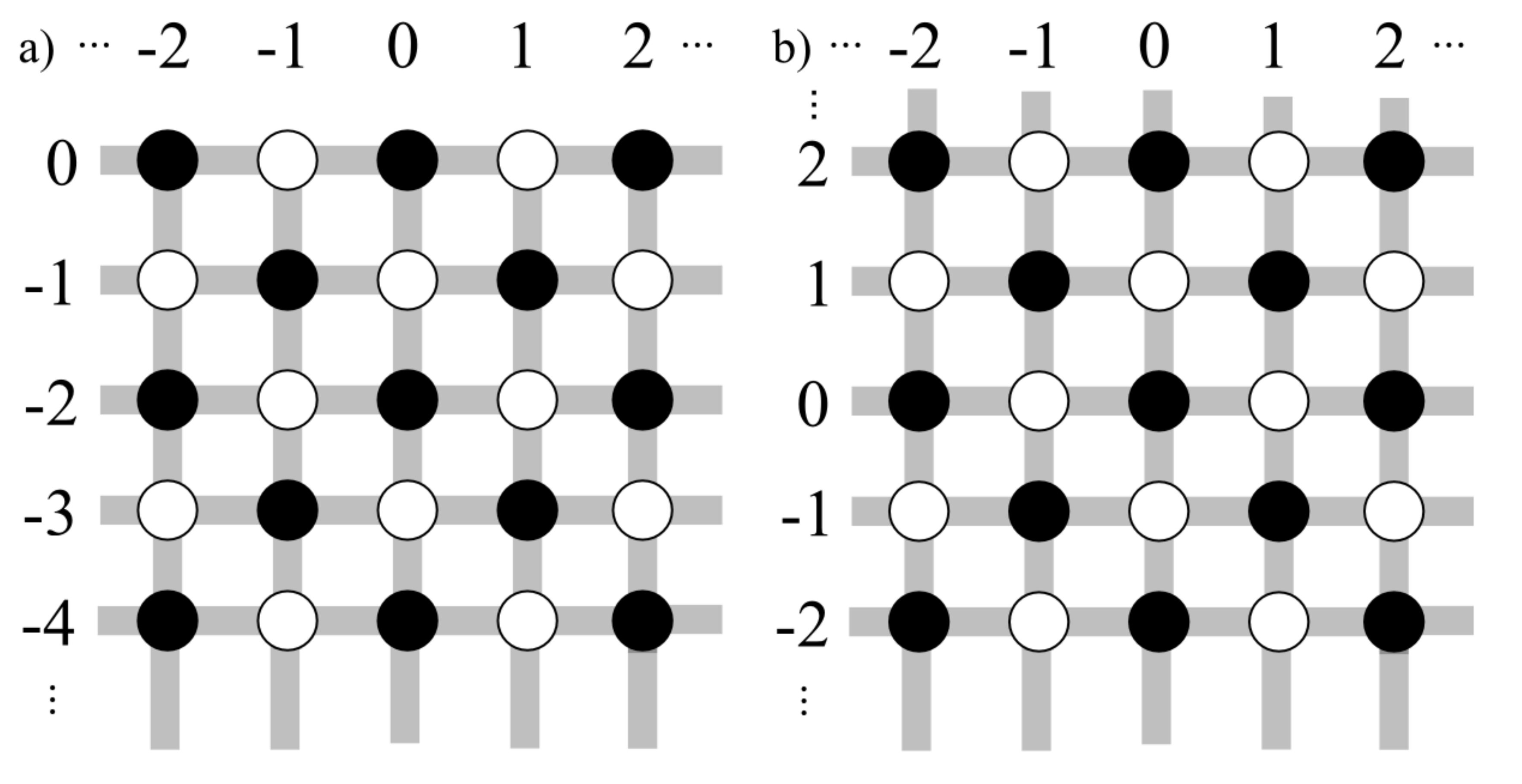}
\caption{Geometries considered in Sec.~\ref{sec:InterDisorder}. a) an ininite half-plane with one edge. b) infinite system. A-(B-) sublattice is denoted by empty (full) sites.}
\label{fig:HalfPlane}
\end{figure}

Here, the system geometry will be an infinite (lower) half-plane, with only one hard edge (located at $y=0$), see Fig.~\ref{fig:HalfPlane}a. We will now take advantage of the tunnelling probability mapping possible between non-interacting and strongly interacting regimes for the classical system, Eq.~(\ref{eqn:Map}). By rescaling the tunnelling probabilities, we obtain the dynamics of a single particle  strongly interacting with a randomly and uniformly distributed sea of strongly interacting particles with a finite density $\nu=N/L^2$, which in essence is a mean-field method.

We track the trajectories of one particle that is initially located on the edge, or $y_0$ sites below the edge, in either sublattice $A$ (which is an edge mode for $\nu=0$ and $y_0=0$) or sublattice $B$ (which is a bulk mode for $\nu=0$). After $m$ periods ($4m$ time steps), we measure the horizontal position where the particle has landed, averaged over many realisations, and obtain its mean speed and trajectory. These quantities are of interest since they are dramatically affected by strong interactions, even for the ideal case of $P_t=1$ -- for which this model is exact also in the quantum case -- as soon as we deviate slightly from the non-interacting (or zero density) case. Brute force Monte Carlo simulations of these phenomena are remarkably slow, as the variances of the distributions grow in time. We have therefore taken a different route to calculate particle trajectories and speeds. As can be seen in Fig.~\ref{fig:HalfPlane}, for each time period there are 6 different types of sites, corresponding to $y_0=0$, $1$ and $\ge 2$, for sublattices $A$ and $B$. Any site further away than $1$ site from the edge will not feel it after one period. We then calculate the one-period transition matrix $M(\mathbf{i},\mathbf{j})$ for each type of site, which is further explained in Appendix~\ref{sec:AppTransfer}, and the time-dependent probabilities $P(\mathbf{i};t)$ are calculated as
\begin{equation}
P(\mathbf{i};t)=\sum_{\mathbf{j}}M(\mathbf{i},\mathbf{j}) P(\mathbf{i};t-T).\label{transfer}
\end{equation}  
The average positions $\langle \mathbf{i} \rangle_t$ and speeds $\mathbf{v}\equiv \langle \mathbf{i} \rangle_{t+T}-\langle \mathbf{i} \rangle_t$ of the particle as a function of time immediately follow from Eq.~(\ref{transfer}).

\begin{figure}[h]
\includegraphics[width=0.49\textwidth]{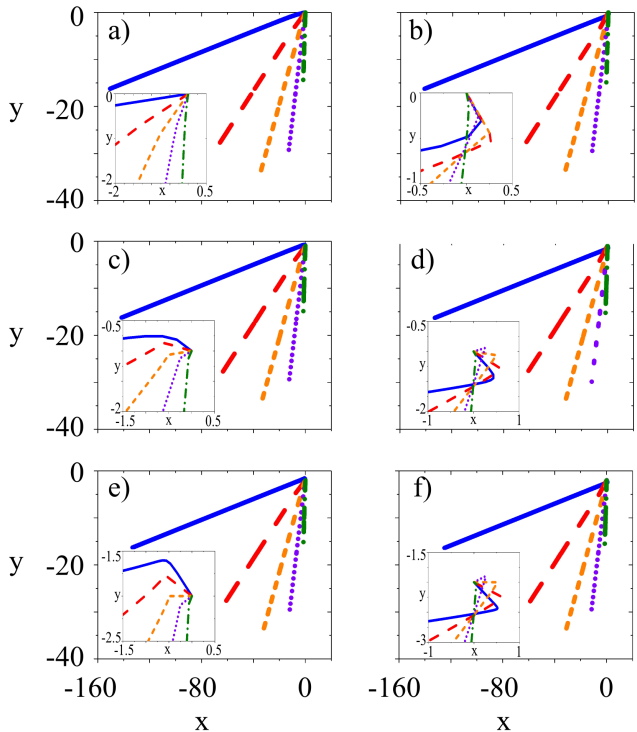}
\caption{Trajectories of the average position over 2000 time periods starting from, a) A site row 0, b) A site row 1, c) A site row 3, d) B site row 0, e) B site row 1 and f) B site row 2. From each starting point we consider 5 fillings, solid (blue) $\nu = 0.1$, dashed (red) $\nu = 0.3$, short dash (orange) $\nu = 0.5$, dot (purple) $\nu = 0.7$ and dash dot (green) $\nu = 0.9$. The inset in each figure gives a close up look at the short time trajectories, of order 10-100 periods.}
\label{fig:Trajectories}
\end{figure}

It is easy to see that interactions play the role of Markovian disorder: at each time step, an impurity will be located in the neighbouring site, say $\mathbf{j}$, of the tracked particle's position with probability $\nu$. However, the next time the particle is neighbouring the site $\mathbf{j}$, the impurity will be present with probability $\nu$. That is, the system has lost memory since impurities are allowed to move around in the lattice. Of course the impurities move in a defined way due to the driving, Fig.~\ref{fig:Model}a, and for high $\nu$ and $P_t \approx 1$ the background will retain memory over short times, of the order of one period. However, over longer times the background's memory is repeatedly washed out by the many interactions present in the system. This is to be contrasted with static, diagonal disorder, for which the first time the particle neighbours $\mathbf{j}$, an impurity will be present with probability $\nu$, but if that is realised, and the particle ends up neighbouring site $\mathbf{j}$ in subsequent time steps, an impurity will be located there with unit probability. This corresponds to a complete memory. An interesting possible extension to this could be the study of the robustness of topology in the driven model, as in certain one-dimensional models, Markovian disorder is known to be able to destroy the topological properties of the system much easier than static disorder \cite{Obuse2011}.

\begin{figure}[t]
\includegraphics[width=0.45\textwidth]{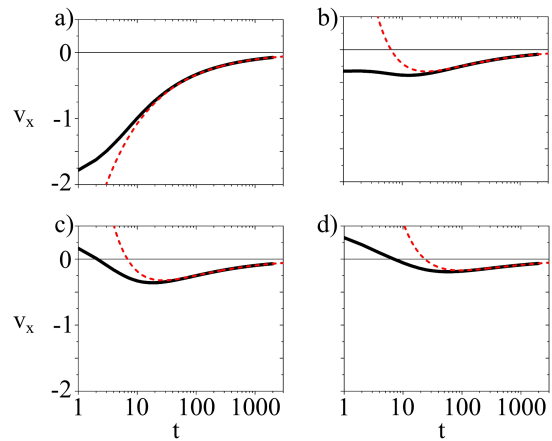}
\caption{The average velocity ($v_x$) in $x$ direction as a function of the number of periods ($t$), with fits (red dash) from Eq.~(\ref{eqn:Fit}). Starting in a) A site row 0, b) A site row 1, c) B site row 0, and d) B site row 1. The red dashed line is the fir from Eq.~(\ref{eqn:Fit}).}
\label{fig:xVelocities}
\end{figure}

The average trajectories of the tracked particle initially on and near the edge are plotted in Fig.~\ref{fig:Trajectories}, for different filling factors. The variance in the displacement is large, it is of the order of the displacement, due to the nature of the problem. We plot the average trajectory over many realizations, and there are many other paths possible over the long time scale considered. In the short-distance (or short time) limit, the trajectories are more intricate and do depend on the initial sublattice. As is seen in the inset of Fig.~\ref{fig:Trajectories}, for any $\nu\ne 0$, the particle's average trajectory is initially already moving towards the left (as expected for edge modes) and downwards for initial $A$-sublattice while for initial $B$-sublattice the particle begins by moving towards the right (as expected for bulk modes), but corrects its trajectory shortly thereafter and continues to behave as an edge mode. This effect is actually quite robust near the edge. To see this, we have estimated the average long-time speed in the $x$-direction as a function of time by fitting its asymptotic behaviour, for a fixed value of $\nu=0.1$, and a variety of starting distances from the edge. The results, together with their corresponding fits to the function
\begin{equation}
v_x(t)=a+bt^{-1/2}+ct^{-1}
\label{eqn:Fit}
\end{equation}
are plotted in Fig.~\ref{fig:xVelocities}. The resulting fits are compatible with $a\equiv 0$ in Eq.~(\ref{eqn:Fit}), which corresponds to $\langle v_x \rangle_t\to 0$ at infinite time, but show a very slow deceleration of the particle as $dv_x/dt \propto t^{-3/2}$ at long times. This can be interpreted as a very slow decay of the edge states that have been coupled over the run time into bulk states, which nevertheless will stop moving at sufficiently long times.

We now consider a geometry which is infinite in all directions, i.e. no edge, as shown in Fig.~\ref{fig:HalfPlane}b, at a low filling $\nu = 0.1$. We launch a particle and again evolve through 2000 periods for starting on an A and B site. Fitting to Eq.~(\ref{eqn:Fit}), our results are again compatible with $\langle v_x \rangle_t\to 0$ at infinite time. This is what would be expected for the non-interacting case because the particle will be in a cyclotron motion. By looking at the trajectories, however, it can be seen that there is a small drift from the cyclotron motion expected for the particle. In 2000 periods the particle on average has moved half a lattice spacing in x and y in the directions of the bulk mode. This is the effect of the interactions present in the system, however the mean-field sea of particles is so dilute that we observe only these small drifts away from the cyclotron motion expected for no interactions.

\section{Quantum Versus Classical Dynamics}
\label{sec:QuantClass}

\begin{figure}[t]
\includegraphics[width=0.45\textwidth]{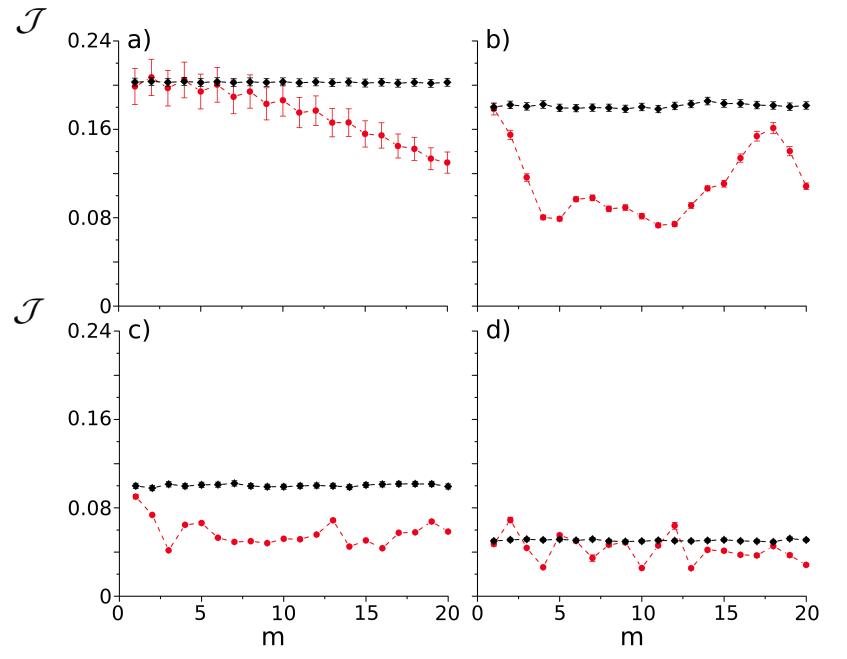}
\caption{The average directional edge current $\mathcal{J}$ as a function of number of periods, m, for 4 particles in 16 lattice sites ($\nu = 0.25$). Circles (red) give spinless fermions and diamonds (black) give classical hardcore particles, for classical probabilities a) $J = 1.539 \: (P_t = 0.999)$, b) $J = 1.249 \: (P_t = 0.9)$, c) $J = 0.785 \: (P_t = 0.5)$, d) $ J = 0.524 \: (P_t = 0.25)$.}
\label{fig:CurrentPeriod}
\end{figure}

In this section we compare simulations for the classical case to a quantum case of spinless fermions. Hardcore bosons and spinless fermions are well known to have an exact mapping in one dimension \cite{Girardeau1960}, but this is not valid in higher dimensions \cite{Crepin2011,Rigol2005} due to their different statistics. Our consideration of hardcore interactions would be the classical limit, as we consider it, for both hardcore bosons and spinless fermions. In this section for the quantum case we will consider spinless fermions, which are hardcore but have the advantage of being non-interacting.

\begin{figure}[t]
\includegraphics[width=0.45\textwidth]{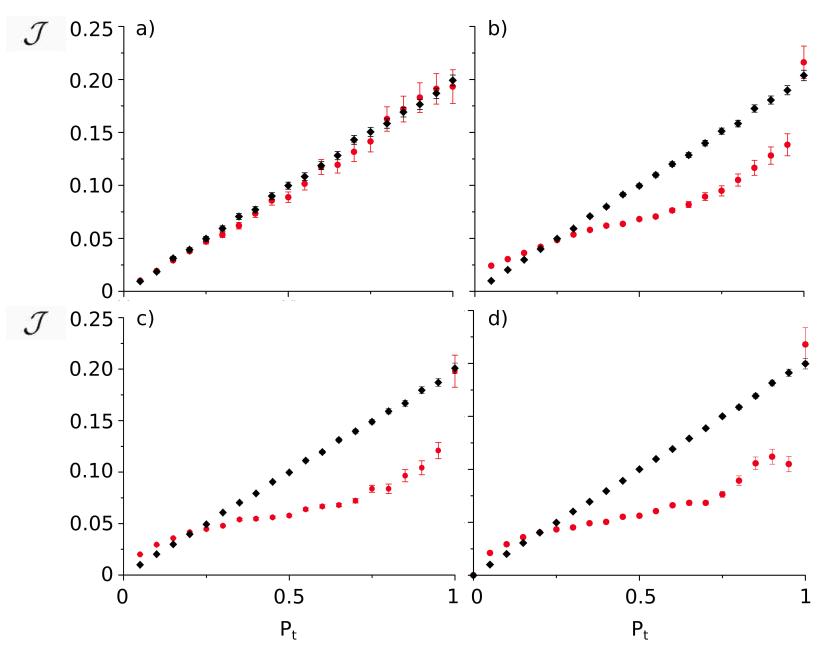}
\caption{The average directional edge current $\mathcal{J}$ as a function of the classical probability, $P_t$, for 4 particles in 16 lattice sites. Circles (red) give spinless fermions and diamonds (black) give classical hardcore particles, for number of periods a) $m = 1$, b) $m = 5$, c) $m = 10$, d) $m = 20$.}
\label{fig:CurrentProbRange}
\end{figure}

The simulations for the quantum case are conducted by evolving the density matrix of the system,
\begin{equation}
\rho (t_m) = e^{-i H(t_m)} \rho(t_{m-1}) e^{i H(t_m)},
\label{eqn:Evolution}
\end{equation}
where the density matrix is given by,
\begin{equation}
\rho(t_m) = \mid \Psi(t_m) \rangle \langle \Psi(t_m) \mid .
\end{equation}
The state $\mid \! \! \Psi(t_m) \rangle$ is the spinless fermion many-body wavefunction at time $t_m$. We calculate the initial density matrix, $\rho(t_0)$, then evolve for each time step Eq.~(\ref{eqn:Evolution}). During the first and third steps of the driving, Fig.~\ref{fig:Model}a, the directional edge current, Eq.~(\ref{eqn:GeneralCurrentEdge}), is calculated. We record the edge current for each period and the sum over all periods. The initial positions of the particles are randomly distributed and the state evolved for many realizations to obtain the average directional edge current.

\begin{figure}[t]
\includegraphics[width=0.45\textwidth]{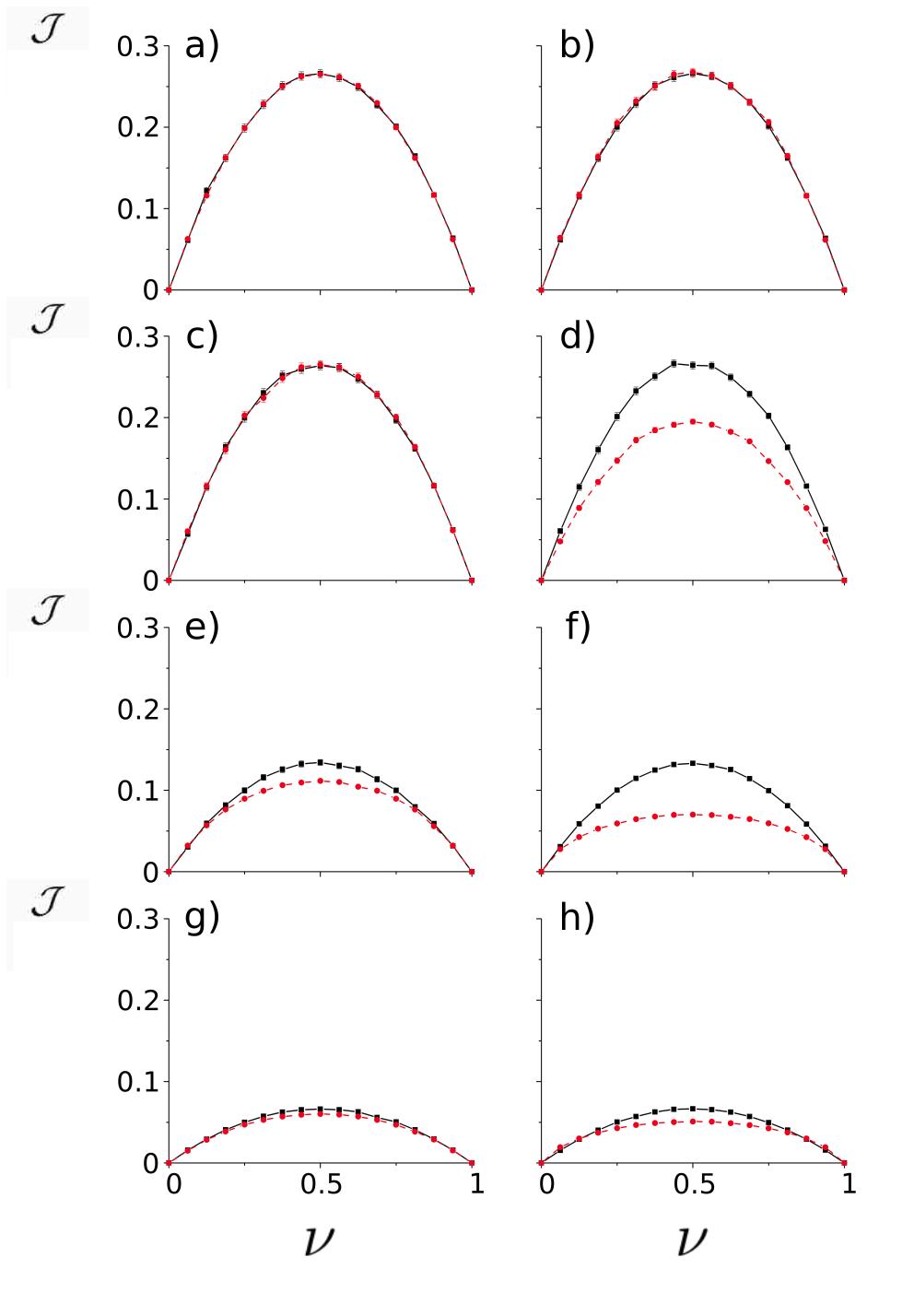}
\caption{The average directional edge current $\mathcal{J}$ as a function of the filling, $\nu$, of a 16 site lattice. Circles (red) give spinless fermions and squares (black) give classical hardcore particles, for tunnelling coefficients of a-b) $J = \pi / 2 \: (P_t = 1)$, c-d) $J \approx 1.539 \: (P_t = 0.999)$, e-f) $J \approx 0.785 \: (P_t = 0.5)$, and g-h) $J \approx 0.524 \: (P_t = 0.25)$, after $m=1$ (a, c, e, g) and $m=30$ (b, d, f, h) time periods of evolution.}
\label{fig:ExactComparison}
\end{figure}

We first consider the directional edge currents dynamical properties, where we record the directional edge current for each time period of evolution. In Fig.~\ref{fig:CurrentPeriod}, we give the average directional edge current over many random initial configurations of 4 particles in 16 lattices sites. We compare the classical hardcore particles and the quantum spinless fermions. Plotting this quantity allows us to observe the effect of the growth of interference for quantum particles compared to the classical case for a range of $J$ and $P_t$, relation between these given by Eq.~(\ref{eqn:ProbRelation}). We evolve the system up to 20 time periods, which is a substantial propagation on a small lattice. Note, in the classical simulations of Sec.~\ref{sec:Sim} where a 100x100 lattice was used we evolve through 100 time periods. For Fig.~\ref{fig:CurrentPeriod}a we have set $J = 0.98 \times \pi/2 \: (P_t = 0.999)$, we observe that to the order of 10 periods the classical and quantum system are in agreement. This is considerable compared to the lattice size and confirms the original thoughts in Sec.~\ref{sec:Classical}, that for values of $P_t \sim 1$ and short times the classical probabilities are a good approximation. However, Fig.~\ref{fig:CurrentPeriod}b $J = 1.249 \: (P_t = 0.9)$, shows that this region of agreement is small, with deviations becoming large after one time period. As one would expect the quantum and classical cases do not agree for $J = 0.785 \: (P_t = 0.5)$, Fig.~\ref{fig:CurrentPeriod}c, where the contribution of interference and possible superposition are at their peak. When the tunnelling probability is small, $P_t = 0.25$ Fig.~\ref{fig:CurrentPeriod}d, we observe substantial agreement between classical and quantum models for long times relative to the lattice size. This is not surprising considering that the contribution of superposition states and interference diminishes as $P_t (J) \rightarrow 0$.

\begin{figure}[t]
\includegraphics[width=0.465\textwidth]{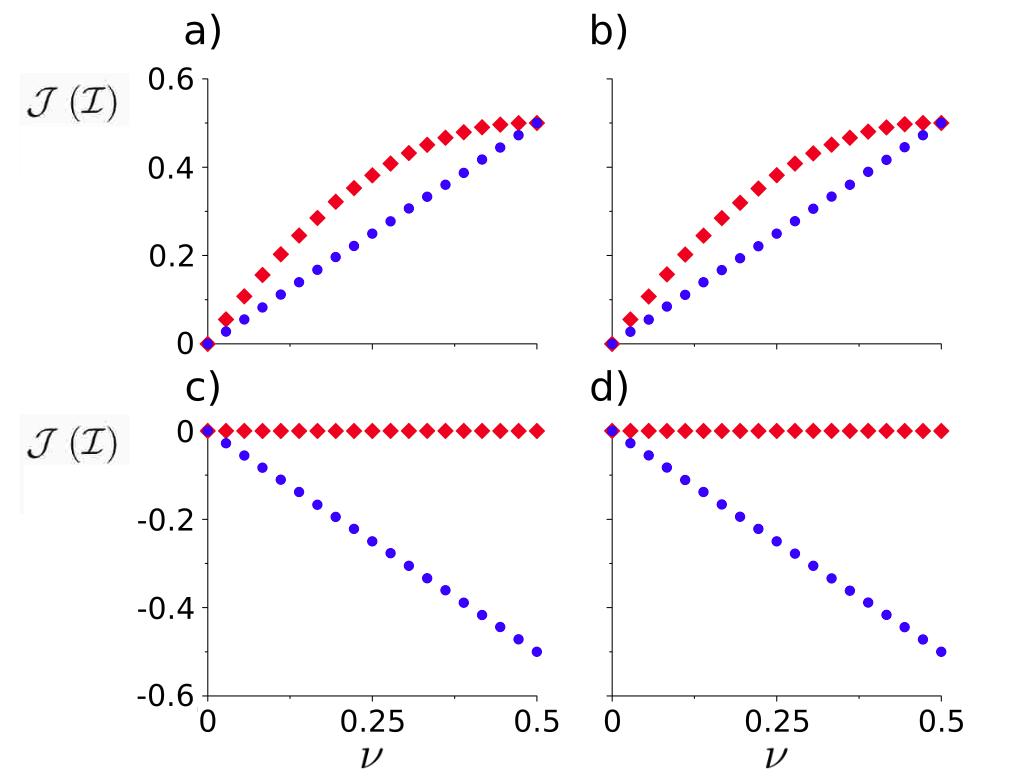}
\caption{Average edge current $\mathcal{I}$ and directional edge current $\mathcal{J}$ on a 36 site lattice, with tunnelling strength $J=\pi/2$ ($P_t = 1$) and $m=30$ time evolutions. Diamonds (red) denote $\mathcal{J}$ and circles (blue) denote $\mathcal{I}$. a) classical case with initial occupation of A sites only, b) quantum case with initial occupation of A sites only, c) classical case with initial occupation of B sites only, and d) quantum case with initial occupation of B sites only.}
\label{fig:ParticleRangeAandBP1}
\end{figure}

In Fig.~\ref{fig:CurrentProbRange} we investigate the directional edge currents relationship to the tunnelling probability $P_t$, over a number of period evolutions. As expected at the special point $P_t = 1 (J=\pi/2)$, the classical and quantum cases are in perfect agreement. However for small deviations from $P_t = 1$, we observe substantial discrepancy between classical and quantum models, in agreement with Fig.~\ref{fig:CurrentPeriod}. However, for all time evolutions conducted we observe an approximate agreement for small $P_t$. The largest discrepancies between the classical and quantum regimes are interestingly when $P_t \rightarrow 1$. This is where interference and superposition states in the system quickly propagate from the initial state, seen in Fig.~\ref{fig:CurrentPeriod}b and Fig.~\ref{fig:CurrentProbRange}b and c. For very small deviations from $P_t = 1$, of order $10^{-3}$, we observe agreement between quantum and classical dynamics on the small lattice, Fig.~\ref{fig:CurrentPeriod}a. The propagation of superposition states and interference is slower for small $P_t (J)$, and hence the small discrepancies between classical and quantum results observed in this region.

To further investigate, we compare the quantum and classical particles in the 16 lattice sites over all $\nu$, for various tunnelling coefficients with short and long time evolution in Fig.~\ref{fig:ExactComparison}. As we have already seen in all previous results, the quantum and classical directional edge currents are exactly equivalent in directional edge current at the special point, $J = \pi / 2 \: (P_t = 1)$, for short and long timescales. As we change the tunnelling coefficient we observe agreement between the cases for one time period of evolution, with the exception of the centre of the filling range for $J \approx 0.785 \: (P_t = 0.5)$, due to small amounts of interference occurring during the time period. For all long time evolutions, except the special point, we see a reduction in the directional edge current of the quantum particle due to interferences, this is the same as we observed in Fig.~\ref{fig:CurrentPeriod}.

\section{Comparison between current and directional current}
\label{sec:ComparCurrent}

So far, we have investigated the directional edge current, as defined by Eq.~(\ref{eqn:GeneralCurrentEdge}). Its behaviour was studied for both the classical and quantum cases in Secs.~\ref{sec:Sim} and \ref{sec:QuantClass} with fully random initial states. The directional edge current observable was required due to the random initial states considered, as discussed in Sec.~\ref{sec:DefinitionCurrent}. This results in the edge current, defined by Eq.~(\ref{eqn:GeneralCurrent}), averaged over all possible initial states, being zero.

\begin{figure}[t]
\includegraphics[width=0.49\textwidth]{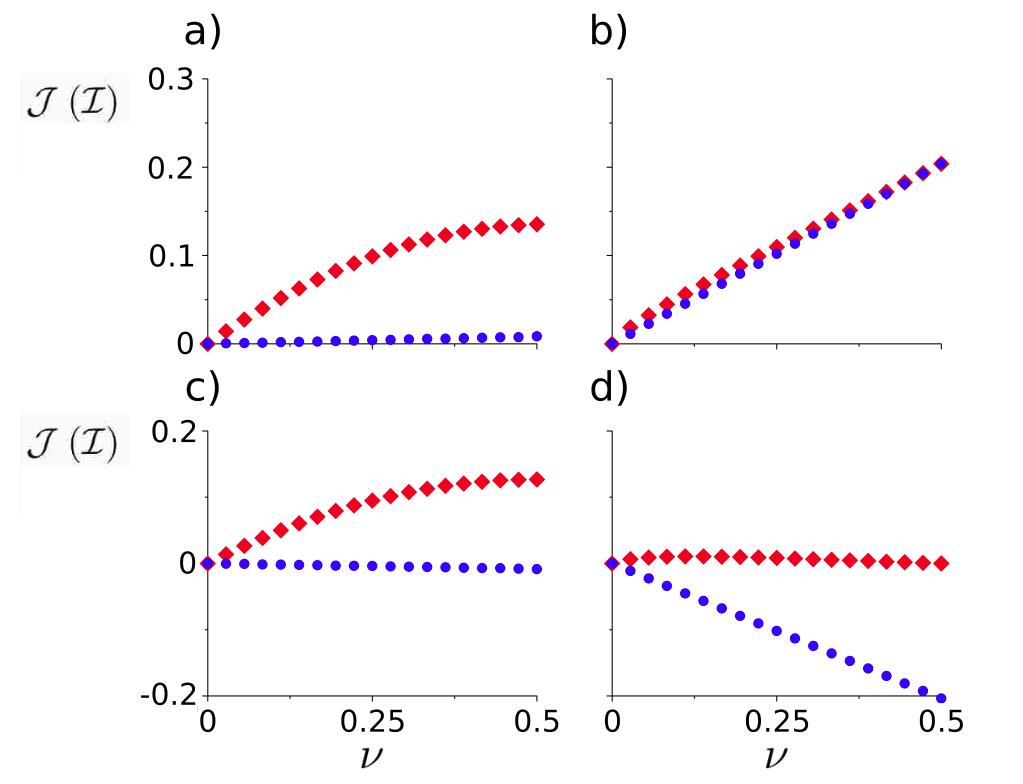}
\caption{Similarly to Fig.~\ref{fig:ParticleRangeAandBP1}, but with a tunnelling strength of $J \approx 0.785$ ($P_t = 0.5$) and $m=30$ time evolutions.}
\label{fig:ParticleRangeAandBP05m30}
\end{figure}

\begin{figure}[t]
\includegraphics[width=0.49\textwidth]{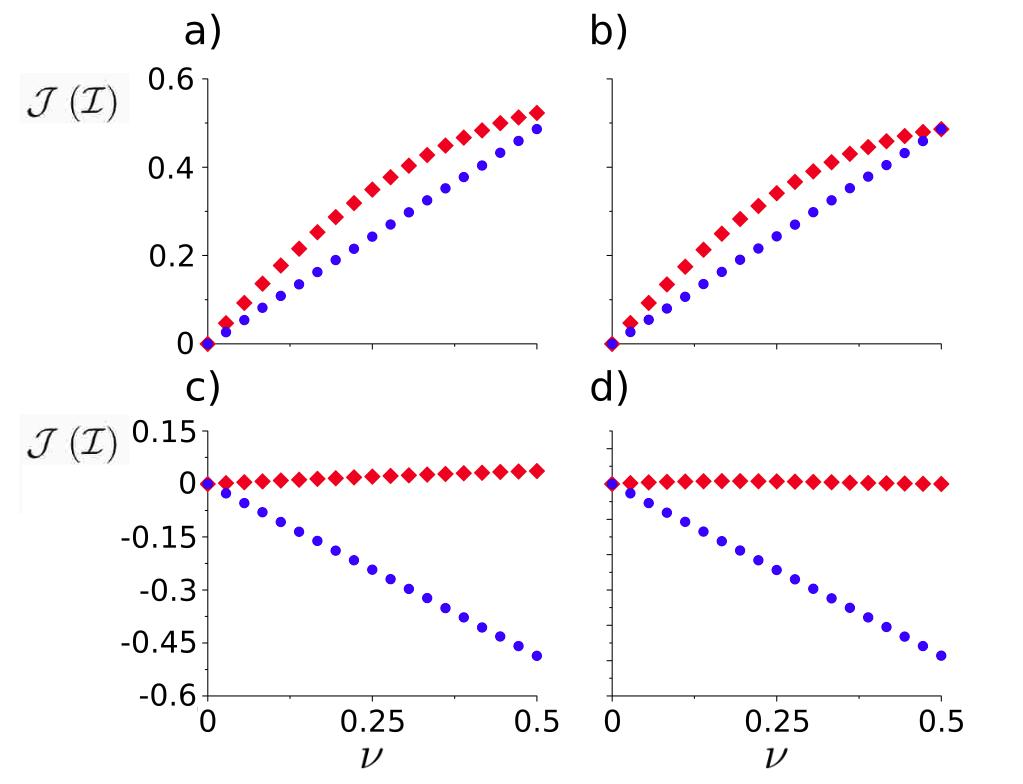}
\caption{Similarly to Fig.~\ref{fig:ParticleRangeAandBP1}, but with a tunnelling strength of $J \approx 1.249$ ($P_t = 0.9$) and $m=30$ time evolutions.}
\label{fig:ParticleRangeAandBP09m30}
\end{figure}

In this section, we consider the relationship between the directional edge current and the edge current, Eqs.~(\ref{eqn:GeneralCurrentEdge}) and (\ref{eqn:GeneralCurrent}) respectively, on the top edge of the system, without loss of generality. Instead of random initial states we will consider initial states that are topologically non-trivial, in the sense that only A or B lattice sites are initially occupied, with the occupation of only A sites resulting in a persistent edge current for $J = \pi / 2$ ($P_t = 1$), as we will discuss below. Throughout this section we consider a system of 36 lattice sites, and for the current observables average over $10^5$ realizations of a given filling.

In the ideal case $J = \pi / 2$ ($P_t = 1$), the dynamics of the system are well prescribed by the driving. As a result, if we start with a state of only A(B) sites being occupied then after any integer number of time periods there will again be only A(B) sites occupied. In the quantum system, with the ideal case, such a topologically non-trivial state would show a persistent, constant current. A particle starting on an A site of the top edge will in each time period move two steps to the left along the edge, hence it follows an edge path and will have a non-zero, positive current. All particles on other A sites of the lattice evolve as a bulk path. The bulk paths of the second row will contribute to the current on the upper edge, as they move in a `cyclotron' motion between the second and first rows, see the driving steps of Fig.~\ref{fig:Model}a. This contribution is negative, hence is expected to result in a decrease of the edge current in comparison to the directional edge current, for which this bulk motion has zero contribution. If only B sites are initially occupied, there are only bulk paths on the top edge, with the edge paths of the bottom row now occupied. We expect the edge current to be negative by considering the motion of the bulk modes on the edge. With each bulk mode of the edge moving one step to the right in the first step of each cycle. This perceived current does not depend on where we define the edge. If we move the boundary of the edge further into the lattice, we will still have a bulk mode contributing to the edge current. Of course, for only B sites initially occupied in the ideal case the directional edge current will be zero. Therefore, we refer to the initial state of the occupation of only A sites as a topologically non-trivial state, whereas the initial state of the occupation of only B sites is topologically trivial (no edge modes are occupied in the ideal case). These observations are, of course, contained in Fig.~\ref{fig:ParticleRangeAandBP1}. We conclude that in the ideal case, the directional edge current is a better observable for the occupation of edge paths, as the full current is negative with the occupation of B sites where no edge paths are occupied.

\begin{figure}[t]
\includegraphics[width=0.49\textwidth]{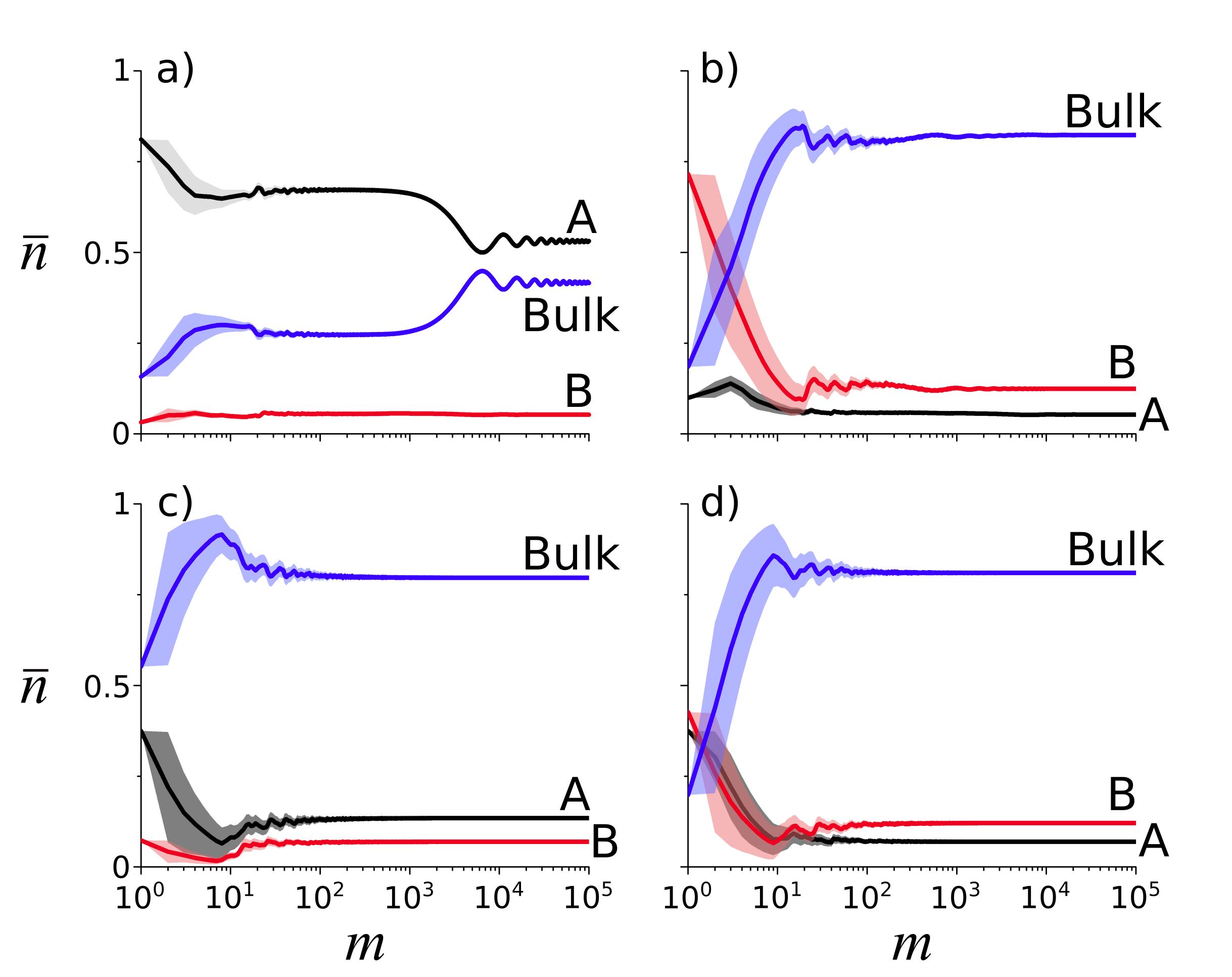}
\caption{Average occupancy $\bar{n}$ in regions of the lattice, of a single-particle for the quantum case in a 36 site lattice after a number of periods, $m$. The lattice is split into three regions, A sites on the edge (black) denoted by `A', B sites on the edge (red) denoted by `B' and all other sites of the lattice (blue) denoted by `Bulk'. The variance is given by the shaded regions of corresponding colour. a) Particle launched on A site on edge at $m=0$ with $J \approx 1.249$, b) Particle launched on B site on edge at $m=0$ with $J \approx 1.249$, c) Particle launched on A site on edge at $m=0$ with $J \approx 0.785$, and d) Particle launched on B site on edge at $m=0$ with $J \approx 0.785$.}
\label{fig:SingleLongTime}
\end{figure}

Away from the ideal case, the edge paths are not robust and can go into the bulk, as was observed by the decrease of current with decreasing tunnelling strength in Secs.~\ref{sec:Sim} and \ref{sec:QuantClass}. Also in Sec.~\ref{sec:InterDisorder}, it was shown that even in the classical ideal case interactions can lead to a particle in an edge path moving into the bulk on average. We now consider tunnelling strengths away from the ideal case, but keep the initial states as the random occupation of either A or B lattice sites. Due to the non-ideal tunnelling strength, we would expect the edge current to go to zero over long time evolutions. As after a large number of periods, as compared to the system size, the lattice will resemble that of random occupation across all sites. Thus, we will recover the result of previous sections, with a zero edge current and a non-zero directional edge current.

In the classical case we observe the tendency of the directional and edge current towards that of the previous sections in Figs.~\ref{fig:ParticleRangeAandBP05m30}a and c for $P_t = 0.5$ and Figs.~\ref{fig:ParticleRangeAandBP09m30}a and c for $P_t = 0.9$, both for $m=30$. As expected the edge current is close to zero, and the directional edge current is of similar form to previous results, e.g. that of Fig.~\ref{fig:Interacting}. We note that due to the non-random initial states the edge current is not exactly zero.

For the quantum case we do not observe the same tendency to previous results, as seen in Figs.~\ref{fig:ParticleRangeAandBP05m30}b and d for $J \approx 0.785$ and Figs.~\ref{fig:ParticleRangeAandBP09m30}b and d for $J \approx 1.249$, both for $m=30$. Instead we observe similar forms to that of the quantum ideal case in Figs.~\ref{fig:ParticleRangeAandBP1}b and d. 

To investigate the non-zero edge currents of the quantum case we consider the dynamics of a single particle initially occupied in an A or B edge site over long time periods. We measure the average over all previous periods of the occupancy, $\bar{n}$, of the particle in A sites on the edge, B sites on the edge and all other sites, which can be denoted as the bulk. Naively we can consider the occupation of A sites on the edge to be the occupation of edge modes in the system, with a higher occupancy corresponding to a larger current. For initial occupancy of an A site on the edge, and $J \approx 1.249$, after long time evolutions we still have the majority of occupation in the A sites of the edge, see Fig.~\ref{fig:SingleLongTime}a. However, with reduced tunnelling strength, $J \approx 0.785$, the particle quickly `decays' into the bulk even at short time scales, see Fig.~\ref{fig:SingleLongTime}c, but there is still a small persistent occupation of A sites on the edge. This A site occupation dominates that of the B sites on the edge, hence the non-zero positive edge current observed with A site occupation in Figs.~\ref{fig:ParticleRangeAandBP05m30} and~\ref{fig:ParticleRangeAandBP09m30}. We observe a similar scenario for the initial occupation of a B edge site. For both tunnelling strengths considered there is a fast transfer of the occupation into the bulk, see Figs.~\ref{fig:SingleLongTime}b and d. However, there is a small persistent occupation of B sites on the edge, which is larger than that of the A sites, resulting in a non-zero negative edge current for B site occupation as seen in Figs~\ref{fig:ParticleRangeAandBP05m30} and~\ref{fig:ParticleRangeAandBP09m30}.

We find that the directional edge current provides a good binary measure of whether there are particles undergoing movement on edge paths or not. Across all conditions considered, we observe negative edge currents if only B sites are initially occupied. Therefore, the combination of $\mathcal{J}$ and $\mathcal{I}$ is necessary for a full understanding of the system.

\section{Conclusions}

In this paper we have considered the edge currents generated by a periodically driven model for many particles for two cases of interactions between particles, non-interacting or strongly interacting (hardcore), in the classical limit. It has been shown that the interactions introduced strongly affect the directional edge current, and also that in this classical system there is no overall flow on the edge, for random initial configurations. In the strongly interacting regime it is easy to see a mean field equivalent, which is of a similar form to mean field traffic models. Using the mean-field argument we can investigate the dynamics of the classical system, where we find trajectories that hint at interesting short time movement with a long time decay of the particles into the bulk. The interactions considered can also be seen as a type of Markovian disorder, where the impurities in the lattice are allowed to move. This opens up the possibility to extend the classical theory developed to be a good approximation beyond $P_t = 1,0$.

We compare the classical and quantum regimes, and the two current observables defined in Sec.~\ref{sec:DefinitionCurrent} for small lattices in the final sections of this work. The classical point-hardcore constraint is replaced in the quantum case by the use of spinless fermions. We confirm the perfect agreement of the classical and quantum models at the special point $P_t = 1$ at all times, and for evolution of the state of only one time period. For low tunnelling probabilities, less than $P_t \sim 0.25$, we find good agreement over substantial evolution times, due to quantum interference playing a smaller role in this region of $P_t$. It is also found that for small deviations from the special point at $J = \pi / 2 \: (P_t = 1)$, the discrepancy between the classical and quantum directional edge current quickly grows with time evolution. From comparing the current and directional current observables along the edge of small lattices, we conclude that a combination of both $\mathcal{J}$ and $\mathcal{I}$ is required for a full understanding of the dynamics of the systems considered.

\section*{Acknowledgements}

We thank A. Spracklen, N. Westerberg and B. Braunecker for useful discussions. C.W.D. acknowledges studentship funding from EPSRC CM-CDT Grant No. EP/L015110/1. P.\"O. and M.V. acknowledge support from EPSRC EP/M024636/1.

\appendix
\section{Calculation of the Path Dependent Currents}
\label{sec:AppPaths}

Here we give the total current contributed by each path, which we label by $\mathit{j}$, that can occur which includes a motion of the particle along the edge, as given in Fig.~\ref{fig:SinglePart}. For the non-interacting (NI) case we have,

\begin{widetext}
\begin{equation}
\begin{aligned}
& \mathit{j}_a^{NI} = & L \nu P_t^2 \qquad \qquad \quad \mbox{ , } & \mathit{j}_b^{NI} = & -L \nu P_t^2 \left(1-P_t\right)^2\mbox{ , } &  \mathit{j}_c^{NI} = & \frac{L}{2} \nu P_t \left(1-P_t\right)^2\mbox{ ,}\\
& \mathit{j}_d^{NI} = & -\frac{L}{2} \nu P_t \left(1-P_t\right)^3\mbox{ , } & \mathit{j}_e^{NI} = & -\frac{L}{2} \nu P_t \left(1-P_t\right)^2\mbox{ , } & \mathit{j}_f^{NI} = & \frac{L}{2} \nu P_t \left(1-P_t\right)\mbox{ ,} \\
& \mathit{j}_g^{NI} = & -\frac{L}{2} \nu P_t^3 \left(1-P_t\right)\mbox{ , } & \mathit{j}_h^{NI} = & -\frac{L}{2} \nu P_t^2 \left(1-P_t\right)^2\mbox{ , } & \mathit{j}_i^{NI} = & \frac{L}{2} \nu P_t^2 \left(1-P_t\right)\mbox{ ,} \\
& \mathit{j}_j^{NI} = & -\frac{L}{2} \nu P_t^2 \left(1-P_t\right)^2\mbox{ , } & \mathit{j}_k^{NI} = & -L \nu P_t^3 \left(1-P_t\right)\mbox{ , } & \mathit{j}_l^{NI} = & -\frac{L}{2} \nu P_t^3 \left(1-P_t\right)\mbox{ ,} \\
& \mathit{j}_m^{NI} = & -\frac{L}{2} \nu P_t^2 \left(1-P_t\right)^2\mbox{ , } & \mathit{j}_n^{NI} = & -\frac{L}{2} \nu P_t^4 \qquad \qquad \mbox{ , } & \mathit{j}_o^{NI} = & -\frac{L}{2} \nu P_t^4 \qquad \qquad \mbox{ ,} \\
& \mathit{j}_p^{NI} = & -\frac{L}{2} \nu P_t^3 \left(1-P_t\right)\mbox{ , } & \mathit{j}_q^{NI} = & -\frac{L}{2} \nu P_t^3 \left(1-P_t\right)\mbox{ .} &  \\
\end{aligned}
\label{eqn:PathsNonInter}
\end{equation}
\end{widetext}

For strongly interacting (SI) particles we have,

\begin{widetext}
\begin{equation}
\begin{aligned}
& \mathit{j}_a^{SI} = & L \nu \left(1-\nu\right)^2 P_t^2 \qquad \qquad \qquad \qquad \quad \: \mbox{ , } & \mathit{j}_b^{SI} = & -L \nu \left(1-\nu\right)^2 P_t^2 \left(\nu + \left(1-\nu\right)\left(1-P_t\right)\right)^2\mbox{ ,} \\
& \mathit{j}_c^{SI} = & \frac{L}{2} \nu \left(1-\nu\right) P_t \left(\nu + \left(1-\nu\right)\left(1-P_t\right)\right)^2\mbox{ , } & \mathit{j}_d^{SI} = & -\frac{L}{2} \nu \left(1-\nu\right) P_t \left(\nu + \left(1-\nu\right)\left(1-P_t\right)\right)^3 \quad \mbox{ ,} \\
& \mathit{j}_e^{SI} = & -\frac{L}{2} \nu \left(1-\nu\right) P_t \left(\nu + \left(1-\nu\right)\left(1-P_t\right)\right)^2\mbox{ , } & \mathit{j}_f^{SI} = & \frac{L}{2} \nu \left(1-\nu\right) P_t \left(\nu + \left(1-\nu\right)\left(1-P_t\right)\right) \quad \: \mbox{ ,} \\
& \mathit{j}_g^{SI} = & -\frac{L}{2} \nu \left(1-\nu\right)^3 P_t^3 \left(\nu + \left(1-\nu\right)\left(1-P_t\right)\right)\mbox{ , } & \mathit{j}_h^{SI} = & -\frac{L}{2} \nu \left(1-\nu\right)^2 P_t^2 \left(\nu + \left(1-\nu\right)\left(1-P_t\right)\right)^2\mbox{ ,} \\
& \mathit{j}_i^{SI} = & \frac{L}{2} \nu \left(1-\nu\right)^2 P_t^2 \left(\nu + \left(1-\nu\right)\left(1-P_t\right)\right)\mbox{ , } & \mathit{j}_j^{SNI} = & -\frac{L}{2} \nu \left(1-\nu\right)^2 P_t^2 \left(\nu + \left(1-\nu\right)\left(1-P_t\right)\right)^2\mbox{ ,} \\
& \mathit{j}_k^{NSI} = & -L \nu \left(1-\nu\right)^3 P_t^3 \left(\nu + \left(1-\nu\right)\left(1-P_t\right)\right)\mbox{ , } & \mathit{j}_l^{SI} = & -\frac{L}{2} \nu \left(1-\nu\right)^3 P_t^3 \left(\nu + \left(1-\nu\right)\left(1-P_t\right)\right)\mbox{ ,} \\
& \mathit{j}_m^{SI} = & -\frac{L}{2} \nu \left(1-\nu\right)^2 P_t^2 \left(\nu + \left(1-\nu\right)\left(1-P_t\right)\right)^2\mbox{ , } & \mathit{j}_n^{SI} = & -\frac{L}{2} \nu \left(1-\nu\right)^4 P_t^4 \qquad \qquad \qquad \qquad \quad \: \mbox{ ,} \\
& \mathit{j}_o^{SI} = & -\frac{L}{2} \nu \left(1-\nu\right)^4 P_t^4 \qquad \qquad \qquad \qquad \quad \: \mbox{ , } & \mathit{j}_p^{SI} = & -\frac{L}{2} \nu \left(1-\nu\right)^3 P_t^3 \left(\nu + \left(1-\nu\right)\left(1-P_t\right)\right)\mbox{ ,} \\
& \mathit{j}_q^{SI} = & -\frac{L}{2} \nu \left(1-\nu\right)^3 P_t^3 \left(\nu + \left(1-\nu\right)\left(1-P_t\right)\right)\mbox{ .} & \\
\end{aligned}
\label{eqn:PathsInter}
\end{equation}
\end{widetext}

To obtain the final forms of the directional edge currents, Eqs.~(\ref{eqn:NonInter}) and (\ref{eqn:Inter}), all paths that move in the direction of the directional edge current from Eqs.~(\ref{eqn:PathsNonInter}) and (\ref{eqn:PathsInter}) respectively are summed.

\section{Calculation of the Transition Matrices}
\label{sec:AppTransfer}

\begin{figure}[h]
\centering
\includegraphics[width=0.2\textwidth]{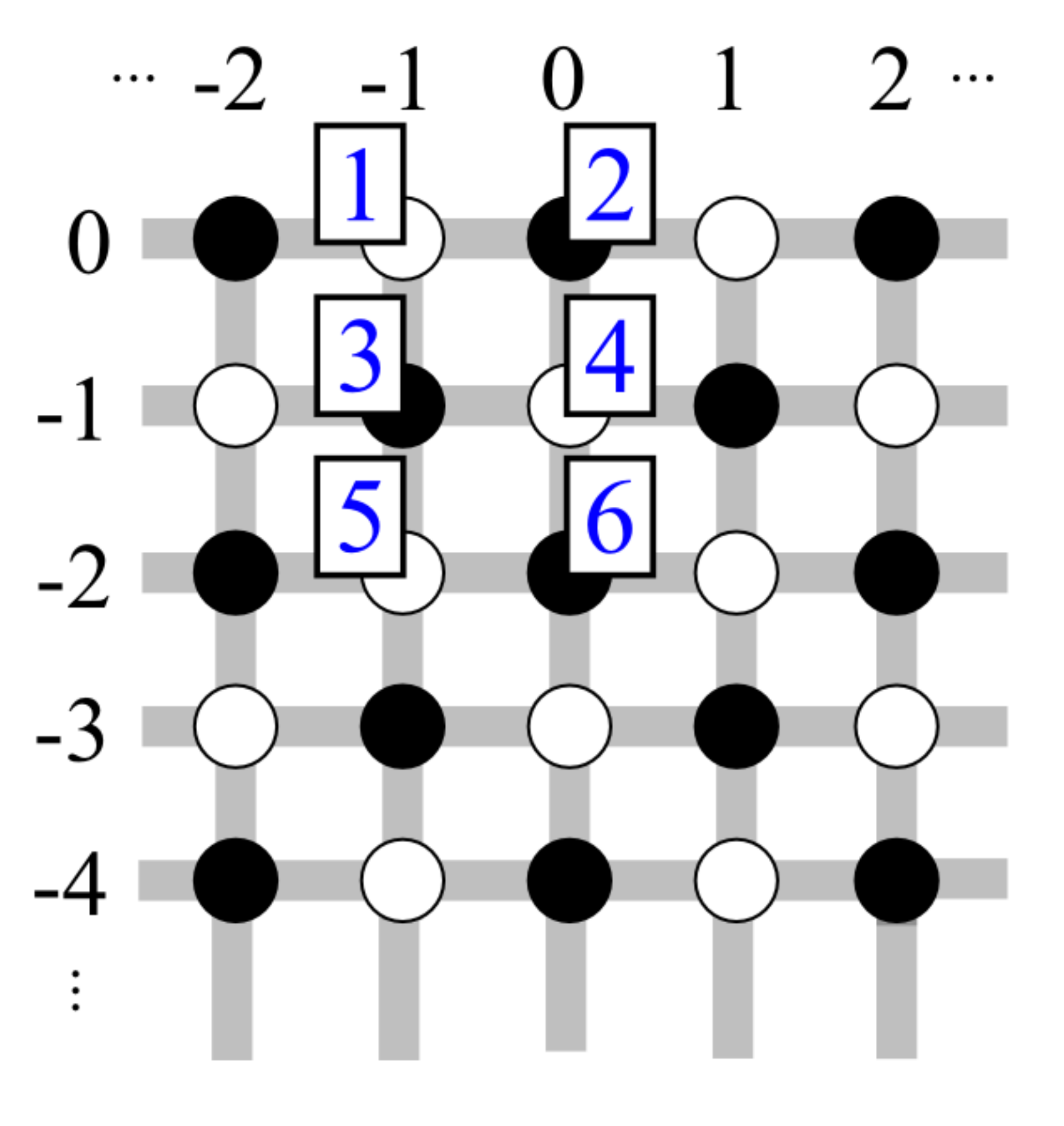}
\caption{The six distinct sites for the half-plane geometry over one time period.}
\label{fig:SixSites}
\end{figure}

To calculate the one-period transition matrix, $M\left(\textbf{i},\textbf{j}\right)$ (Eq.~(\ref{transfer})), for each type of site considered in Sec.~\ref{sec:InterDisorder} requires a more formal approach than the one considered in Sec.~\ref{sec:ApprModel}. For one time period in the half plane geometry of Fig.~\ref{fig:HalfPlane}(a), there are 6 distinct sites, shown in Fig.~\ref{fig:SixSites}. Any site below row -2 is the same as 5 and 6 as the particle can only move a maximum of two spacings in one period. This means once the particle is two rows away from the edge the effect of some sites not being coupled along the edge in time steps 2 and 4 is no longer felt in one period. In the infinite geometry in Fig.~\ref{fig:HalfPlane}(b) there are only two distinct sites given by 5 and 6 in Fig.~\ref{fig:SixSites}.

To find all elements of the transition matrix for one of the distinct sites we simply start at time zero with a single particle occupying the site and then consider for each time step all possible moves. For the first time step this is relatively simple as there are only two possibilities, namely, the particle tunnels with probability $P_t$ or it does not with probability $\left(1-P_t\right)$. We then take these two cases and consider the possible next steps from each. This is repeated for all possibilities from each step. Note that care needs to be taken for time steps 2 and 4 as not all sites are coupled to another site. This method gives a tree which results in the probabilities of transfer to all the possible sites from the starting site. These are then used to construct the transition matrix for that site.

The building blocks of the probabilities of each path are the same as discussed in Sec.~\ref{sec:Contib}. However by using the tree method discussed we ensure no possibility is missed and that the full transition matrix is constructed. For example, for stating at $(\textbf{i}_0,\textbf{j}_0)=(1,0)$ on Fig.~\ref{fig:HalfPlane}(a),
\begin{equation}
\begin{aligned}
M\left(1,0\right) &= (1-P_t)^3 \\ M\left(2,0\right) &= P_t(1-P_t)^3 \\ M\left(2,-1\right) &= P_t^2(1-P_t)^2 \\ M\left(1,-1\right) &= P_t(1-P_t)^3 \\ M\left(1,-2\right) &= P_t^2(1-P_t)^2 \\ M\left(0,-1\right) &= P_t^2(1-P_t)^2 + P_t^2(1-P_t) \\ M\left(0,0\right) &= P_t^3(1-P_t) + P_t(1-P_t)^2 \\ M\left(-1,0\right) &= P_t^2 \\  
\end{aligned}
\end{equation}

\end{document}